\documentclass{aa}  

\usepackage{CJK}
\usepackage{color}
\usepackage{gensymb}
\usepackage{graphicx}
\usepackage{txfonts}
\usepackage{subcaption}         \usepackage{lscape}             \usepackage{placeins}           \usepackage{hyperref}
\usepackage{numprint}
\usepackage[normalem]{ulem}

\providecommand\figref[1]{Fig.~\ref{#1}}
\providecommand\figsref[1]{Figs.~\ref{#1}}
\providecommand\figrefalt[1]{Figure~\ref{#1}}
\providecommand\figsrefalt[1]{Figures~\ref{#1}}

\begin{document}
\begin{CJK*}{UTF8}{gbsn}

   \title{Tidally-induced vertical asymmetries in the LMC bar and disk}

   \author{Shuyu Wang (汪书玉)\inst{1}
        \and Marie Sch\"olch \inst{2,3,4}
        \and Merc\`e Romero-G\'omez \inst{2,3,4}
        \and \'Oscar Jim\'enez-Arranz \inst{5}
        \and Anthony G.A.~Brown \inst{1}
        }

   \institute{Leiden Observatory, Leiden University, Einsteinweg 55, 2333 CC Leiden, The Netherlands\\
        \email{shuyuwang.astro@gmail.com}
        \and Institut de Ci\`encies del Cosmos (ICCUB), Universitat de Barcelona (UB), Mart\'{\i} i Franqu\`es 1, E-08028 Barcelona, Spain
        \and Departament de F\'{\i}sica Qu\`antica i Astrof\'{\i}sica (FQA), Universitat de Barcelona (UB), Mart\'{\i} i Franqu\`es 1, E-08028 Barcelona, Spain
        \and Institut d'Estudis Espacials de Catalunya (IEEC), Esteve Terradas 1, Edifici RDIT, Campus PMT-UPC, 08860 Castelldefels (Barcelona), Spain
        \and Lund Observatory, Division of Astrophysics, Department of Physics, Lund University, Box 43, 22100 Lund, Sweden}

  \abstract
   {Past observations and simulations have shown that the bar in the Large Magellanic Cloud (LMC) may be tilted with respect to the disk plane, and that the outer LMC disk could be warped. These studies propose that the tilted bar and the warp in the LMC are induced by the Small Magellanic Cloud (SMC) pericentric passages. }
   {We analyze and quantify the effect of the SMC passage on the vertical position and kinematic asymmetries of both the bar and disk, as a function of different impact parameters or initial
   conditions.
   } 
   {We used seven representative simulations from the KRATOS suite of $N$-body simulations of the LMC, SMC and Milky Way (MW) system for this analysis. 
   To classify bar particles and analyze the bar and disk separately, we adopted a classification method based on persistent homology. To quantify the vertical asymmetries in the bar and disk of the LMC-like galaxy, we applied Fourier decomposition. We constructed {\it Gaia}~DR3 mock catalogs to study whether the observed vertical asymmetries in {\it Gaia}~DR3 could be caused by astrometric uncertainties.
   }
   {We detect strong $m = 1$ vertical asymmetries in both the bar and disk of the LMC-like galaxy for all interacting simulations after the pericentric passage of the SMC-like galaxy, indicative of a warped bar and disk. We find that the amplitude of the induced $m = 1$ vertical asymmetry strongly correlates with the distance between the two galaxies. We also detect $m = 2$ vertical asymmetries in the bar of the LMC-like galaxy for some simulations after the passage of the SMC-like galaxy, indicative of a buckling bar.  }
   {We conclude that the SMC passages induce vertical asymmetries in both the bar and the disk of the LMC, compatible to those measured in {\it Gaia}~DR3 maps. Tidal interactions naturally give rise to a warped and buckling LMC bar, and a warped LMC disk. Comparison between our results and {\it Gaia}~DR3 observations suggest that models with a higher mass ratio between the SMC and the LMC ($m_\mathrm{SMC} /{M_\mathrm{LMC}} \sim 0.22$) may be favorable for reproducing the observed tilt of the LMC bar, and that the most recent SMC pericentric passage occurred at most $210$ to $300$~Myr ago.}   
   \maketitle
\nolinenumbers

\section{Introduction}
\label{sec:intro}
The Large Magellanic Cloud (LMC) and Small Magellanic Cloud (SMC) are the largest and closest satellite galaxies of the Milky Way (MW). The LMC, a Barred Magellanic Spiral galaxy, exhibits asymmetric morphology. It has an off-centered bar \citep[e.g.,][]{2004ApJ...614L..37Z, 2009ApJ...703L..37S, 2024A&A...688A..51J,2025A&A...701A.227S}, which is likely tilted \citep[e.g.,][]{2000ApJ...545L..35Z,2012MNRAS.421.2109B,2018ApJ...866...90C,2025ApJ...988...79R,2025A&A...698A..88J}, and a warped outer disk \citep[e.g.,][]{vandermarel01,2012MNRAS.421.2109B,2018ApJ...866...90C,2022A&A...666A.103S,2025A&A...698A..88J,2026ApJ..1003....9O}. These asymmetries could be a result of the ongoing interactions between the Magellanic Clouds (MCs), as estimated by the {\it in-plane} kinematic asymmetries in the LMC bar and outer disc \citep{2025A&A...701A.227S}. In particular, the LMC has undergone a close encounter (within $5$~kpc) with the SMC approximately $150-300$\,Myr ago \citep[e.g.,][]{2011ApJ...737...29O,2022ApJ...927..153C, 2022MNRAS.510..445C}.

Regarding the effect of the tidal interactions on the vertical component, we consider the bar and the disk as two separate kinematic components of the LMC. \citet{2018ApJ...866...90C} and \citet{2025A&A...698A..88J} estimate a $5\degree \text{-} 15\degree$ tilt of the LMC bar with respect to the disk plane. In Model 2 of the $N$-body smoothed-particle hydrodynamics (SPH) simulation of the LMC, SMC and MW system in \citet{2012MNRAS.421.2109B}, where a recent direct collision between the MCs occurs, the authors detect a $10\degree \text{-} 15\degree$ tilt in the LMC bar. Recently, \citet{2025ApJ...988...79R} use isodensity ellipse fitting and measure a $\sim 8.6\degree$ tilt at the present time in the LMC bar in Model 2 from the same simulation, consistent with the observations.

On the other hand, several studies indicate that the LMC disk is warped \citep[e.g.,][]{2001AJ....122.1807V,2012MNRAS.421.2109B,2018ApJ...866...90C,2022A&A...666A.103S,2025A&A...698A..88J,2026ApJ..1003....9O}. This fact is supported by both observational studies \citep[e.g.,][]{vandermarel01, 2018ApJ...866...90C}
and by numerical simulations, where \citet{2012MNRAS.421.2109B} find that both the outer stellar and gas disks of the LMC are warped and distorted in Model 2 of their simulations. 
\citet{2025A&A...698A..88J} construct three-dimensional models of the LMC using SDSS and {\it Gaia}~DR3 data, which qualitatively reproduce both the tilted bar and the warped outer disk. Using red clump stars from {\it Gaia}~DR3 data, \citet{2026ApJ..1003....9O} find that the warp in the outer LMC is U-shaped.

In this work, we use seven representative simulations with varying initial conditions from KRATOS \citep{2024A&A...688A..51J}, a suite of $N$-body simulations of the LMC, SMC and MW system system. We search for vertical position and velocity asymmetries in the bar and disk of the simulated LMC-like galaxy, and relate them with the tidal interaction between the SMC-like galaxy and the LMC-like galaxy. To analyze vertical asymmetries in the bar and disk of the simulated LMC-like galaxy separately, we adopt the topology-based bar classification method in \citet{2025CNSNS.14908923S}. We investigate whether vertical asymmetries in the bar and disk are a direct result of interactions between the two galaxies, and how different tidal interaction parameters and initial conditions could affect the amplitudes of the vertical asymmetries. We qualitatively compare the detected vertical asymmetries in the bar and disk of the LMC-like galaxy in KRATOS simulations to those in \citet{2012MNRAS.421.2109B} and observational studies \citep[e.g.,][]{2018ApJ...866...90C,2025A&A...698A..88J}, which allows us to constrain the mass ratio between the two galaxies, and to estimate the time since the last SMC pericentric passage. Finally, we analyze the vertical asymmetries in the kinematic maps of the LMC in {\it Gaia}~DR3, and use {\it Gaia} mock catalogs to provide caveats on how to interpret the results.

This paper is organized as follows. In Sect.~\ref{sec:sim}, we review the main characteristics of the simulations used in this work. In Sect.~\ref{sec:method}, we describe the bar classification method, the centering and aligning strategy and the Fourier decomposition method, which allows us to quantify the vertical asymmetries. We present our main results in Sect.~\ref{sec:result} and Sect.~\ref{sec:all_models}. In Sect.~\ref{sec:discuss}, we compare our results with Model 2 of the numerical simulations in \citet{2012MNRAS.421.2109B}, {\it Gaia}~DR3 observations and {\it Gaia} mock catalogs. We summarize our results in Sect.~\ref{sec:conclude}.

\section{Simulations}
\label{sec:sim}
In this work, we analyzed seven simulations of KRATOS \citep{2024A&A...688A..51J}, which consists of a suite of 28 pure $N$-body simulations that model the evolution of an isolated or interacting LMC-like galaxy, denoted by $G_\mathrm{LMC}$ following \citet{2024A&A...688A..51J}, with varying parameters. These simulations are K5, K6, K10, K11, K12, K20, and K21. Each simulation models an isolated $G_\mathrm{LMC}$ (K10), a $G_\mathrm{LMC}$ interacting with an SMC-like galaxy, $G_\mathrm{SMC}$ ($G_\mathrm{LMC}+G_\mathrm{SMC}$; K5, K11, K20), or a $G_\mathrm{LMC}$ interacting with both a $G_\mathrm{SMC}$ and a MW-like galaxy, $G_\mathrm{MW}$ ($G_\mathrm{LMC}+G_\mathrm{SMC} + G_\mathrm{MW}$; K6, K12, K21). The $G_\mathrm{LMC}$ system is modeled as a stellar disk with an exponential profile embedded in a live Navarro-Frenk-White (NFW, \citealt{1996ApJ...462..563N}) dark matter halo. The $G_\mathrm{SMC}$ system is modeled as a live NFW halo, which comprises both the stellar and the dark matter components. The $G_\mathrm{MW}$ is modeled as a live NFW dark matter halo.

These seven simulations are chosen due to the presence of a consistently strong bar ($\Sigma_2 / \Sigma_0>0.2$, see Fig.~7 of \citealt{2024A&A...688A..51J}) in their $G_\mathrm{LMC}$. They can be grouped into three sets based on their initial conditions: K5 and K6 have lower Toomre parameters ($Q = 1.0$) than the other simulations ($Q = 1.2$); K10, K11, K12, K20, and K21 have less massive $G_\mathrm{LMC}$ dark matter halos ($M_\mathrm{LMC, halo} = 0.8\times 10^{11} M_{\odot}$) than those in K5 and K6 ($M_\mathrm{LMC, halo} = 1.8\times 10^{11} M_{\odot}$); K20 and K21 have less massive $G_\mathrm{LMC}$ stellar disks ($M_\mathrm{LMC, disk} = 3\times 10^{9} M_{\odot}$) than the other simulations ($M_\mathrm{LMC, disk} = 5\times 10^{9} M_{\odot}$). All simulations have the same mass for the $G_\mathrm{SMC}$ ($M_\mathrm{SMC, stellar} = 2.6\times 10^{8} M_{\odot}$ and $M_\mathrm{SMC, halo} = 1.9\times 10^{10} M_{\odot}$) and the $G_\mathrm{MW}$ ($M_\mathrm{MW,halo} = 10^{12} M_{\odot}$). The initial conditions of these simulations are provided in Tables~1 and~2 in \citet{2024A&A...688A..51J}.

The simulations are run for $4.68$~Gyr, with a time cadence of $75$~Myr. Following  \citet{2024A&A...688A..51J}, we set the present time of the simulation as the snapshot $4$~Gyr after the initial conditions, which matches
the number of $G_\mathrm{LMC}$-$G_\mathrm{SMC}$ pericentric passages and qualitatively resembles the morphology of the observed LMC disk (see their Sect.~3.2).

\begin{figure}
  \resizebox{\hsize}{!}{\includegraphics{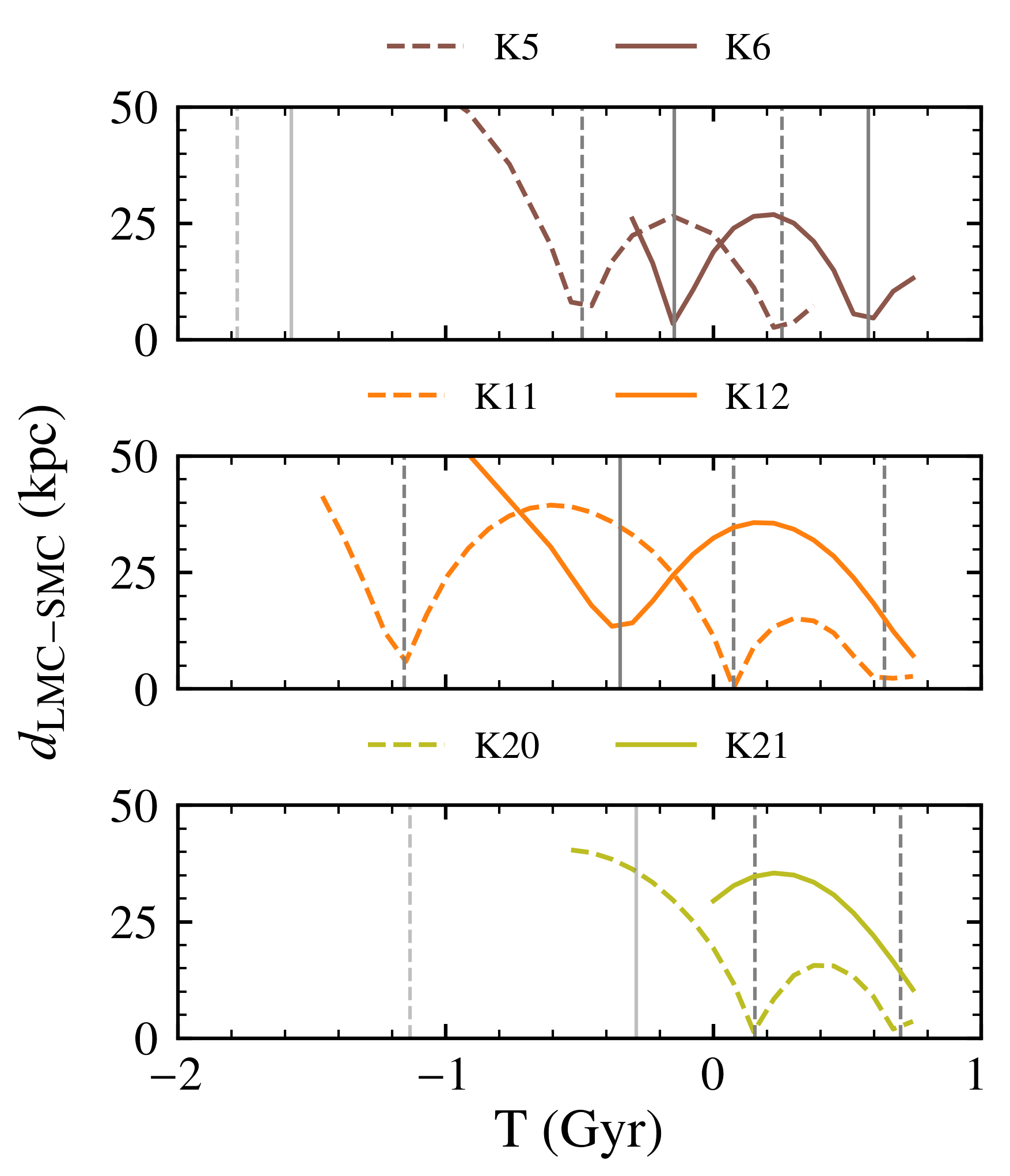}}
  \caption{Time evolution of the distance between the $G_\mathrm{LMC}$ and the $G_\mathrm{SMC}$. The solid (dashed) lines correspond to the $G_\mathrm{LMC}+G_\mathrm{SMC} + G_\mathrm{MW}$ ($G_\mathrm{LMC}+G_\mathrm{SMC}$) models. Colors denote the different sets of simulations, following \citet{2024A&A...688A..51J}. The vertical grey dashed lines correspond to the pericenters of the $G_\mathrm{LMC}+G_\mathrm{SMC}$ models (K5, K11, K20). The vertical grey solid lines correspond to the pericenters of the $G_\mathrm{LMC}+G_\mathrm{SMC} + G_\mathrm{MW}$ models (K6, K12, K21). The pericenters before the full formation of the $G_\mathrm{LMC}$ bar (or the first pericenters for K5 and K6) are denoted by light grey lines.}
  \label{fig:orbit}
\end{figure}

For the interacting models, we provide the time evolution of the distance between the center of mass of the $G_\mathrm{LMC}$ and $G_\mathrm{SMC}$ stars ($d_{\mathrm{LMC-SMC}}$) in \figref{fig:orbit}. The pericentric passage is closer for models with lower $M_\mathrm{LMC}$ that are interacting with only the $G_\mathrm{SMC}$, K11 and K20. Because the first $G_\mathrm{SMC}$ pericentric passage in K5 and K6 is too distant to produce significant effects on the $G_\mathrm{LMC}$, we study snapshots close to and after the second pericentric passage of the $G_\mathrm{SMC}$. For all models, we study the snapshots after the full formation of the $G_\mathrm{LMC}$ bar.

\section{Method}
\label{sec:method}
\subsection{Bar classification}
\label{sec:class}
In order to study the vertical asymmetries in the $G_\mathrm{LMC}$ bar and disk individually, it is crucial to separate the two components. Although methods based on Fourier decomposition \citep[e.g.,][]{1985ApJ...288..438E}, isophotal ellipse fitting \citep[e.g.,][]{1999MNRAS.308..569A}, or machine-learning \citep[e.g.,][]{2018MNRAS.477..894A} have been proposed to classify bars, bar classification remains a non-trivial task. Recently, \citet{2025CNSNS.14908923S} developed a novel, topology-based method for bar classification which only uses the positions and velocities of the star particles in a single snapshot. To identify bar particles in the $G_\mathrm{LMC}$, we adapt their bar classification algorithm and apply it to the three dimensional positions and velocities of the $G_\mathrm{LMC}$ star particles.

We first perform K-means clustering on the absolute Galactocentric radial velocities ($\left|V_R \right|$) of the $G_\mathrm{LMC}$ star particles and divide them into two groups. Three dimensional positions of the star particles belonging to the high absolute radial velocity group form the point cloud in our topological analysis. We perform Vorono\"i tesselation on the point cloud, creating a graph that connects two points with neighboring Vorono\"i cells whose distance is smaller than the maximal distance parameter $d_\mathrm{max}$ (ranges from $0$ to $\infty$) by an edge. For a certain $d_\mathrm{max}$, we identify the component with the largest number of connected points, also known as the largest connected component (LCC). \citet{2025CNSNS.14908923S} tested the number of star particles within the LCC, the number of retrograde star particles within the LCC, and the semimajor axis length $a$ as possible detectors for bar limits (see their Figs.~5,~6 and~10), and determined the number of retrograde star particles within the LCC to be the best one empirically. Therefore, we plot the number of retrograde star particles contained within the LCC versus a range of $d_\mathrm{max}$ values (see \figref{fig:kmeans_dmax_retro}). As $d_\mathrm{max}$ increases, the number of (retrograde) star particles within the corresponding LCC increases steeply, connecting the dense inner region of the galaxy. After the bar is fully included in the LCC, $d_\mathrm{max}$ reaches a limit (corresponding to the knee of the plot), where increasing it further will only include fewer high radial velocity retrograde outliers in the disk region. We define the bar particles as those contained within the LCC corresponding to this limiting $d_\mathrm{max}$. Details of applying this method are provided in App.~\ref{sec:bar_class}. For a full description of the method, we refer the reader to \citet{2025CNSNS.14908923S}.

\begin{figure}
    \centering
  \resizebox{0.85\hsize}{!}{\includegraphics{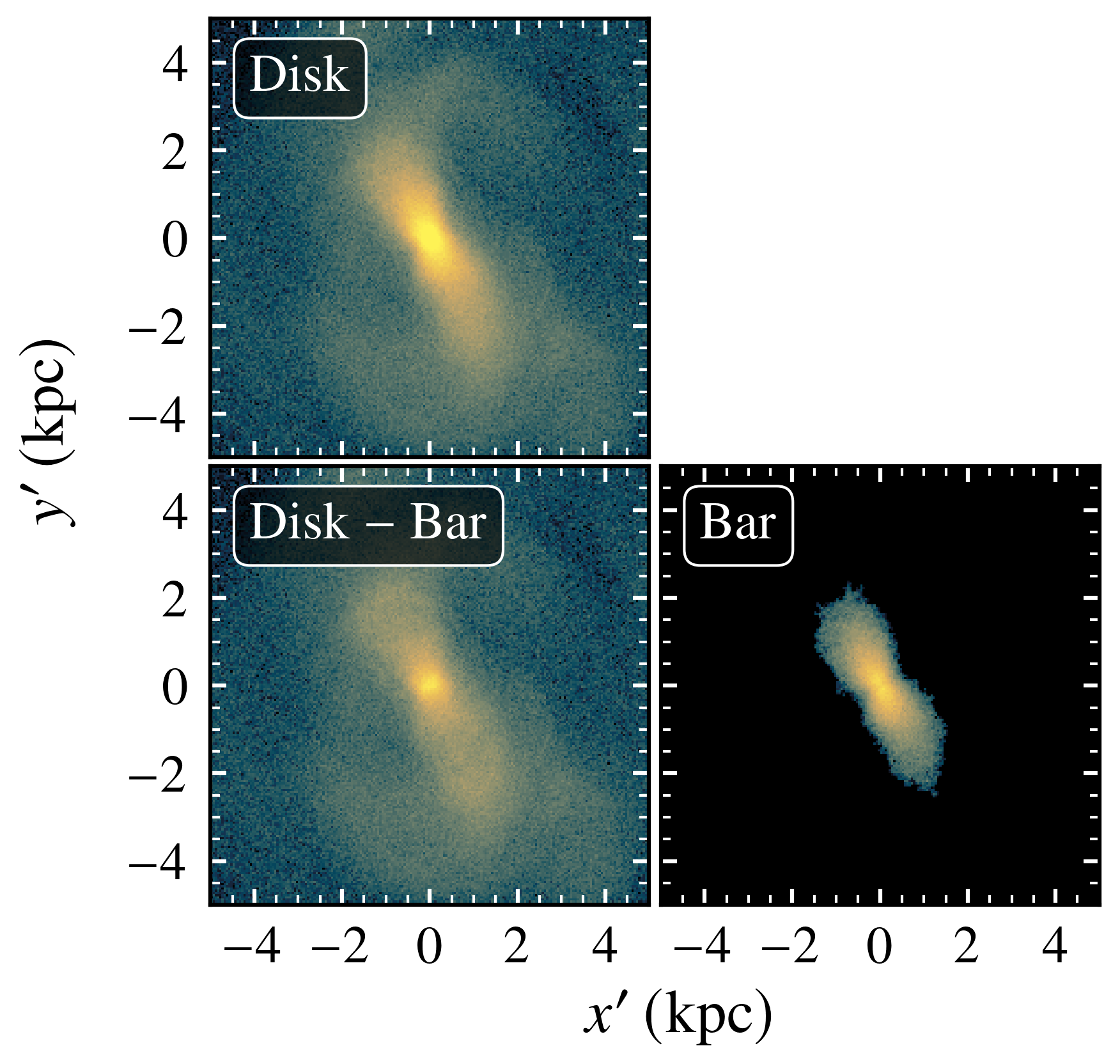}}
  \caption{The face-on log density plot of the $G_\mathrm{LMC}$ (top left), the $G_\mathrm{LMC}$ with bar particles removed (bottom left) and the $G_\mathrm{LMC}$ bar (bottom right) in the K11 simulation at the second $G_\mathrm{SMC}$ passage ($t = 0.075$~Gyr) after applying the bar classification method, centering and aligning the $G_\mathrm{LMC}$ disk.}
  \label{fig:bar_classification}
\end{figure}

As an example, in \figref{fig:bar_classification}, we show the face-on density plot of the $G_\mathrm{LMC}$ (top left), the $G_\mathrm{LMC}$ with bar particles removed (bottom left) and the extracted $G_\mathrm{LMC}$ bar (bottom right) in the K11 simulation at the second $G_\mathrm{SMC}$ passage ($t = 0.075$~Gyr) after applying the bar classification method, centering and aligning the $G_\mathrm{LMC}$ disk (details will be discussed in the next subsection). In the bottom left panel of \figref{fig:bar_classification}, some residual particles remain within the central region of the $G_\mathrm{LMC}$ after removing the bar particles, suggesting that our method prioritizes purity over completeness. \figrefalt{fig:bar_classification} demonstrates the effectiveness of our classification method.

Finally, to check the performance of the method implemented in this work, we also tested a recent star-by-star bar classification method from \citet{2026MNRAS.548ag611I}, which is based on the failure of conservation of the total action for bar particles. We determined that the topology-based method is preferable for this study, considering the relatively low cadence of the KRATOS simulations ($\Delta t = 75$~Myr) compared to that in \citet{2026MNRAS.548ag611I}, where $\Delta t = 10$~Myr. Additionally, in our interacting models, the $G_\mathrm{LMC}$ disk could be severely disrupted by the $G_\mathrm{SMC}$ passages, making it difficult to distinguish between bar and disk particles based solely on the conservation of total action. The implementation of the action-based bar classification method and the differences between the two methods are discussed in App.~\ref{sec:bar_class2}.

\subsection{Centering and aligning the $G_\mathrm{LMC}$}
\label{sec:cen_align}
Here, we present the centering and alignment method for the $G_\mathrm{LMC}$ used for the rest of the work. We first center the reference frame at the density center of the $G_\mathrm{LMC}$ bar following \citet{2024A&A...688A..51J}. We perform principal component analysis (PCA) on the three dimensional positions of the $G_\mathrm{LMC}$ star particles in the region within $R = 5$~kpc using \code{scikit-learn} \citep{scikit-learn}. The normal vector of this region is the smallest principal component. We align the $z$-axis of the reference frame with the normal vector, requiring that the vertical component of the total angular momentum vector of all the $G_\mathrm{LMC}$ star particles has positive $z$. We obtain bar particles using the topology-based bar classification method (see Sect.~\ref{sec:class}). Then we center the $G_\mathrm{LMC}$ disk particles (bar particles removed) at their center of mass (COM).  
We refine our alignment by aligning the reference frame with the re-centered disk particles within $5$~kpc. Henceforth, we denote the $G_\mathrm{LMC}$ in-plane Cartesian coordinates as $(x^\prime,y^\prime,z^\prime)$. In this coordinate system, the $G_\mathrm{LMC}$ rotates counterclockwise.

\begin{figure}
  \resizebox{\hsize}{!}{\includegraphics{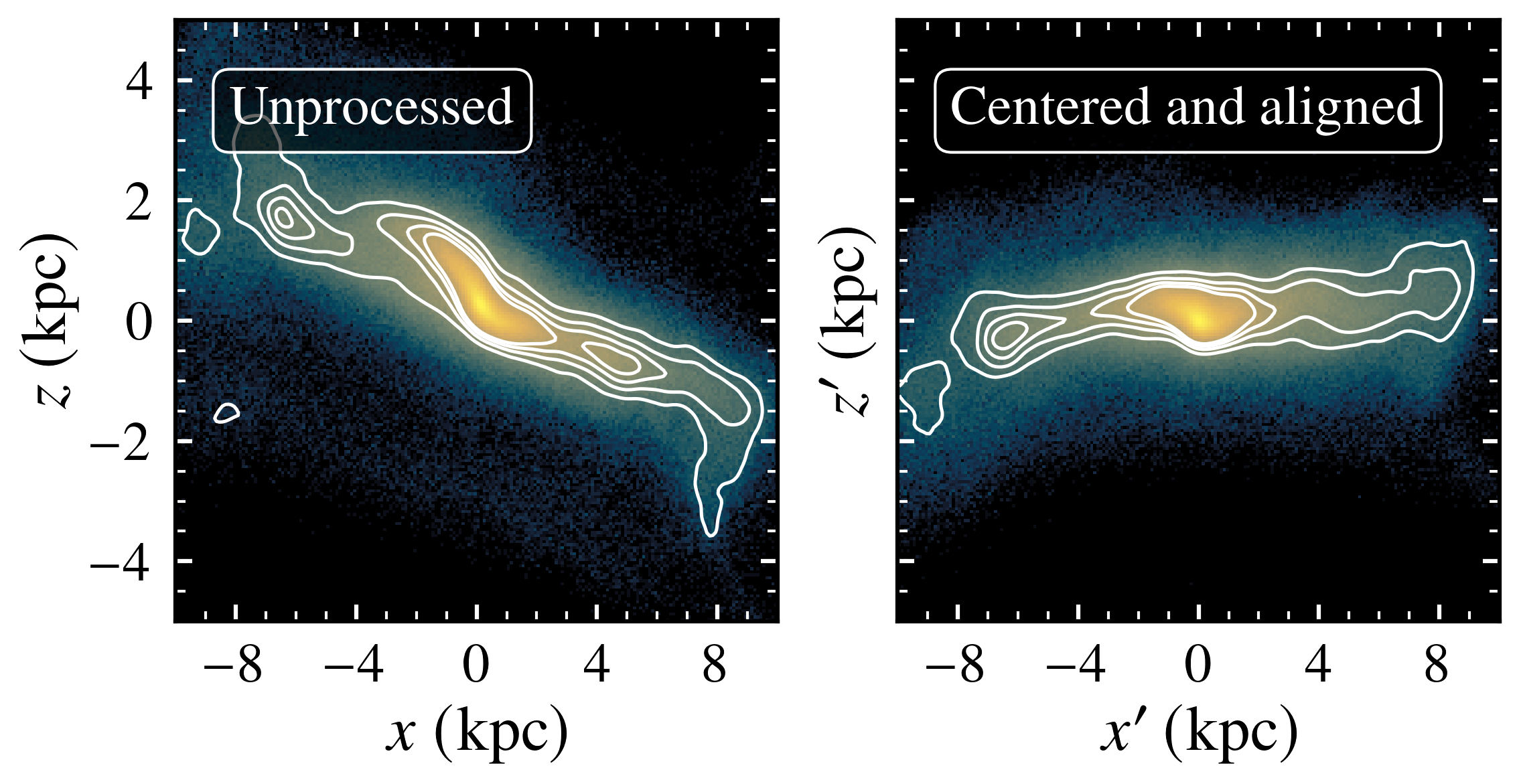}}
  \caption{The unprocessed side-on density plot of the $G_\mathrm{LMC}$ in the K11 simulation at $t = 0.075$~Gyr (left), and the density plot after centering and aligning the $G_\mathrm{LMC}$ (right). The contour lines highlight the overdensities in the $G_\mathrm{LMC}$.}
  \label{fig:center_align}
\end{figure}

\figrefalt{fig:center_align} illustrates the $G_\mathrm{LMC}$ before and after centering and aligning in the K11 simulation at $t = 0.075$~Gyr. The left panel shows the unprocessed snapshot, 
the right panel shows the snapshot after centering at the density center of the bar, aligning with the region within $R = 5$~kpc, and finally aligning with the $G_\mathrm{LMC}$ disk (bar particles removed, centered at the COM of the disk) within $5$~kpc.

\begin{figure}
  \resizebox{\hsize}{!}{\includegraphics{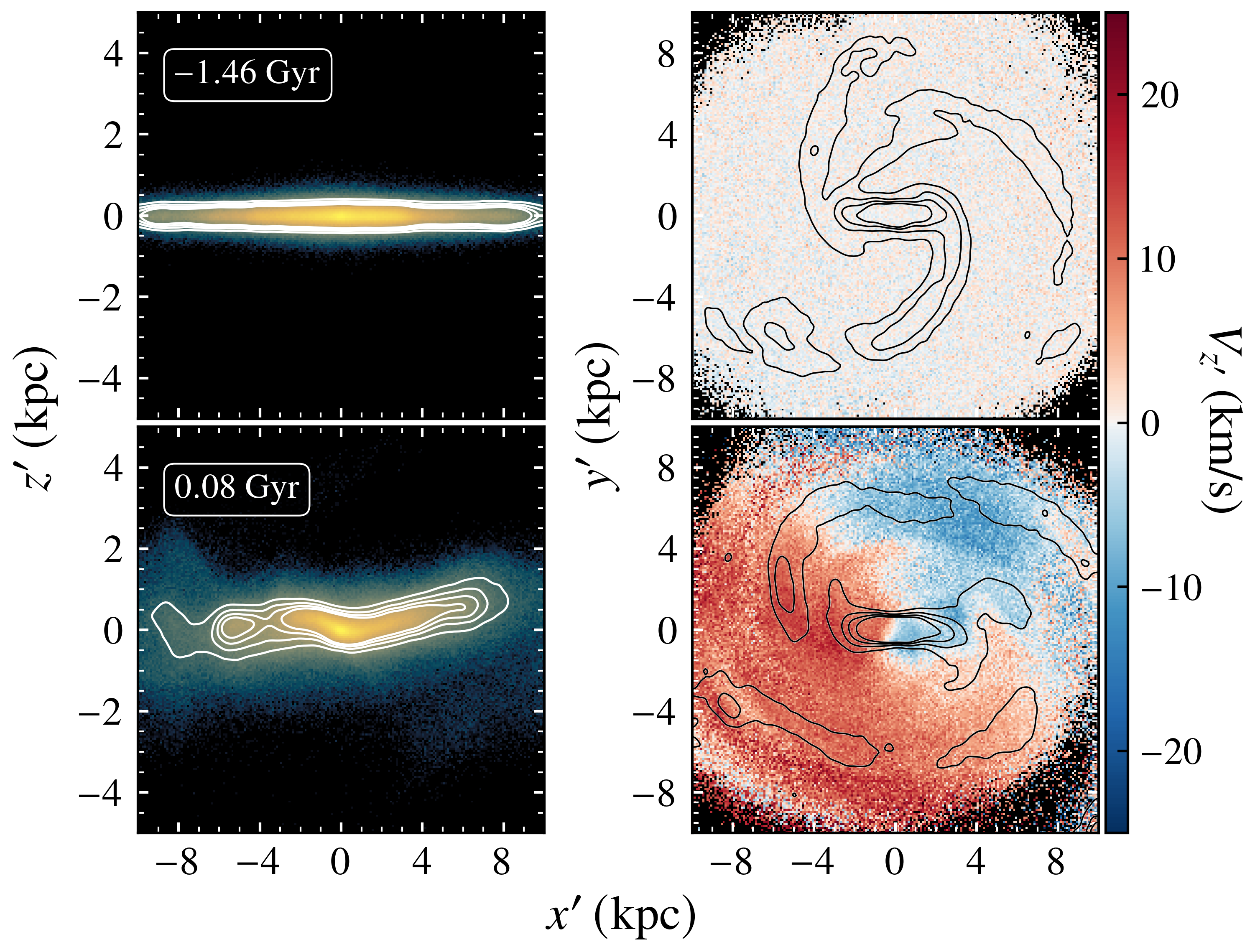}}
  \caption{The centered and aligned $G_\mathrm{LMC}$ at $t = -1.46$~Gyr (before the first $G_\mathrm{SMC}$ passage, top panels) and at $t = 0.075$~Gyr (at the second $G_\mathrm{SMC}$ passage, bottom panels), respectively, in the K11 simulation in the side-on view, color-coded by density (left panels) and in the face-on mean vertical velocity plot (right panels) after applying the bar classification method, centering and aligning the $G_\mathrm{LMC}$ disk. For illustration purposes, we align the $x^{\prime}$-axis with the semi-major axis of the bar.}
  \label{fig:face_side_on_7550_8550}
\end{figure}

\figrefalt{fig:face_side_on_7550_8550} shows the centered and aligned $G_\mathrm{LMC}$ at $t = -1.46$~Gyr (before the first $G_\mathrm{SMC}$ passage, top panels) and at $t = 0.075$~Gyr (at the second $G_\mathrm{SMC}$ passage, bottom panels), respectively, in the K11 simulation in the side-on view, color-coded by density (left panels) and in the face-on mean vertical velocity plot (right panels). The plot demonstrates that before the first $G_\mathrm{SMC}$ passage, the $G_\mathrm{LMC}$ disk is flat and the reference frame aligns well with the disk plane. However, the $G_\mathrm{SMC}$ passage violently disrupts the $G_\mathrm{LMC}$ disk, making it difficult to determine a single $G_\mathrm{LMC}$ disk plane. 

Centering and aligning the $G_\mathrm{LMC}$ disk is a crucial step for the following morphological and kinematic analysis. The analysis of a highly-disrupted disk might suffer from systematics due to not having a fully aligned disk. Therefore, we have tested an alternative alignment method of the $G_\mathrm{LMC}$ disk where we iteratively obtain bar particles after aligning the $G_\mathrm{LMC}$ and center consistently on the density center of the $G_\mathrm{LMC}$. We find no significant differences between the results obtained using the two methods (see App.~\ref{sec:ori_align}).

\subsection{Fourier decomposition}
To quantitatively study the asymmetry in the vertical position, $z^{\prime}$, and vertical velocity, $V_z^{\prime}$, of the $G_\mathrm{LMC}$ bar and disk, we use Fourier decomposition. The vertical asymmetry of a Fourier mode $m$ has amplitude
\begin{align}
    A_{m, z^{\prime}} = \frac{1}{N} \left| \sum_j z_j^{\prime} \mathrm{e}^{im\phi_j} \right| \,, \quad\text{or}\quad    A_{m, V_{z^{\prime}}} = \frac{1}{N} \left| \sum_j V_{{z^{\prime},j}} \mathrm{e}^{im\phi_j} \right| \,,
\end{align}
where $z_j^{\prime}$, $V_{{z^{\prime},j}}$ and $\phi_j$ are the vertical position, velocity and azimuth of a particle $j$, respectively. We associate the $m = 1$ vertical asymmetry for positions with the warp, and the $m = 2$ asymmetry in the bar with bar buckling \citep[e.g.,][]{2006ApJ...645..209D}.

\section{Case study with K10, K11 and K12}
\label{sec:result}
In this section, we show the detected vertical asymmetries in the $G_\mathrm{LMC}$ bar and disk. As an example, we present the morphological analysis and Fourier decomposition results for a set of simulations K10 (isolated $G_\mathrm{LMC}$), K11 ($G_\mathrm{LMC}+G_\mathrm{SMC}$) and K12 ($G_\mathrm{LMC}+G_\mathrm{SMC} + G_\mathrm{MW}$).

\subsection{Time evolution of the $G_\mathrm{LMC}$ morphology and kinematics}
Pre-processing the simulations (Sect.\ref{sec:method}) allows us to track the morphological and kinematic evolution of the $G_\mathrm{LMC}$ bar and disk. 
The figures below illustrate the complex morphological and kinematic evolution of the $G_\mathrm{LMC}$ when interacting with the $G_\mathrm{SMC}$. In particular, vertical asymmetries arise in the $G_\mathrm{LMC}$ after each $G_\mathrm{SMC}$ passage.

\begin{figure*}
    \centering
\includegraphics[width=0.48\textwidth]{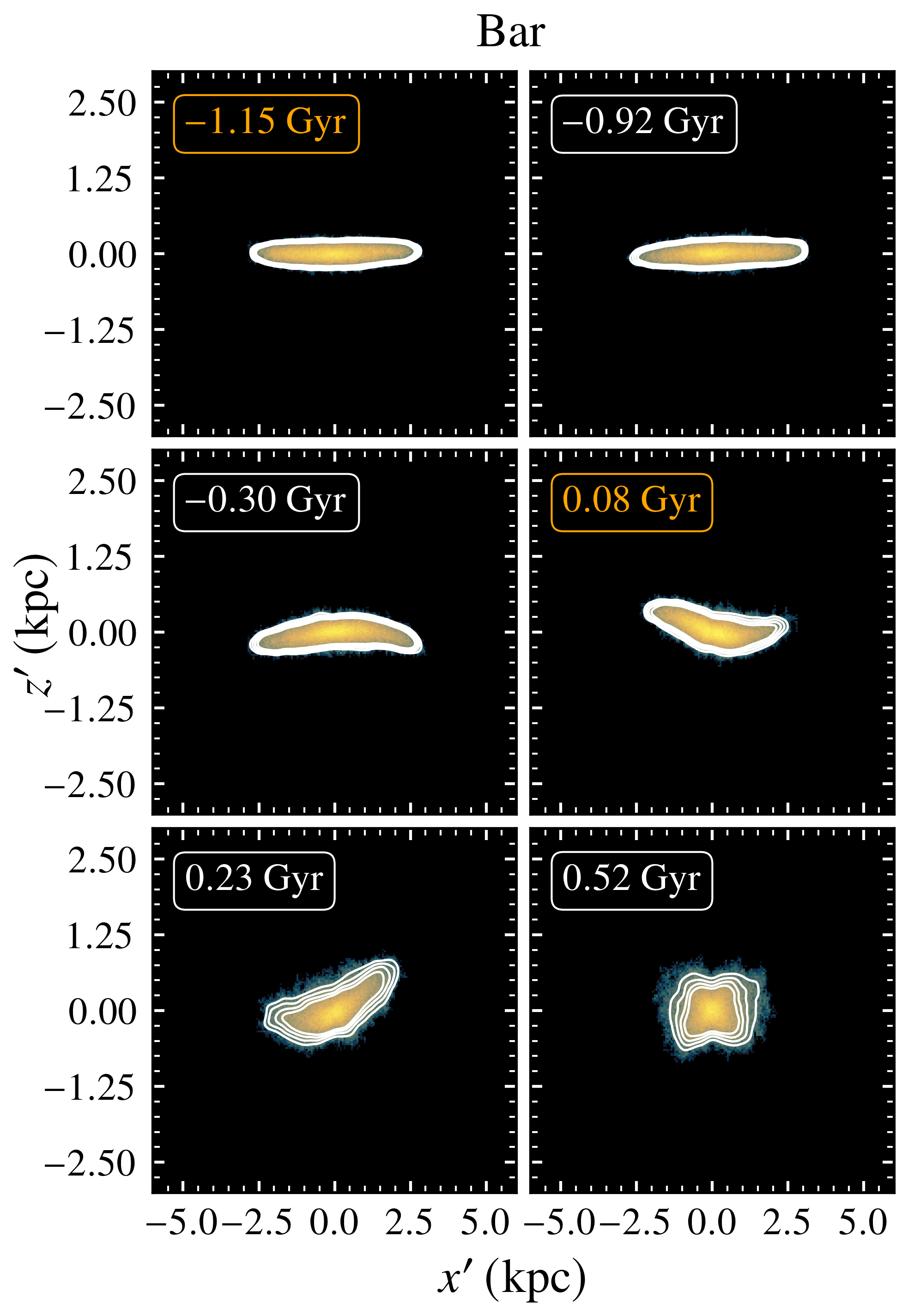}
    \hspace{0.3cm}
\includegraphics[width=0.48\textwidth]{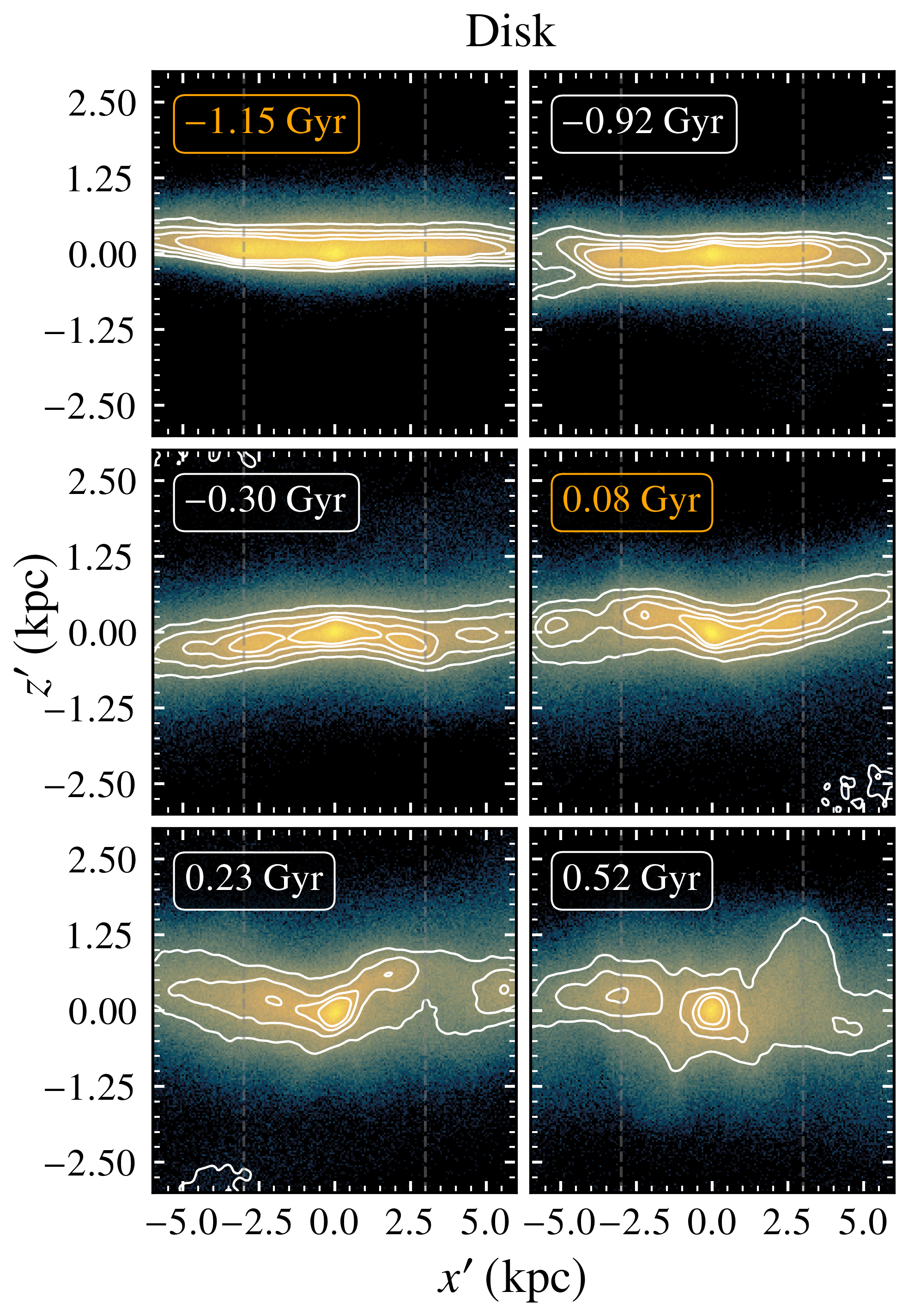}
    \caption{Time evolution of the side-on density distribution of the $G_\mathrm{LMC}$ bar (left panels) and disk (right panels) in the K11 simulation. For illustration purposes, the $x^{\prime}$-axis is aligned with the semi-major axis of the bar. The text in the upper left corner denotes the time of the snapshot, while the orange text highlights the time at the first and second pericentric passages of $G_\mathrm{SMC}$. The grey dashed line denotes the radius cut for the $G_\mathrm{LMC}$ disk region, $3\leq R \leq 6$~kpc.}
  \label{fig:K11_bar_disk}
\end{figure*}

In the left panels of \figref{fig:K11_bar_disk}, we show the side-on density distribution of the $G_\mathrm{LMC}$ bar in selected snapshots of K11. The text in the upper left corner of each panel denotes the time of the simulation, while the orange text highlights the time of the pericentric passage of the $G_\mathrm{SMC}$. The panels illustrate that at the first $G_\mathrm{SMC}$ passage, the $G_\mathrm{LMC}$ bar remains relatively flat (first and second panels, from top left to bottom right). The $G_\mathrm{LMC}$ bar buckles shortly afterwards, characterized by the downward curve (banana shape, third panel). The second pericentric passage induces an additional warp to the already buckling bar (characterized by the upward curve, fourth and fifth panels, also see \figref{fig:buckling_tilting_model}), and the bar appears lopsided. The effect of applying an additional $m = 1$ vertical asymmetry to a buckling bar is demonstrated by a simple model in App.~\ref{sec:buckling_model}. In the end, a vertically symmetric boxy bulge emerges at the center of the $G_\mathrm{LMC}$ (sixth panel).

The right panels of \figref{fig:K11_bar_disk} shows the time evolution of the side-on density distribution of the $G_\mathrm{LMC}$ disk (bar particles removed, $3\leq R \leq 6$~kpc) in the K11 simulation. After the first $G_\mathrm{SMC}$ passage, the outer $G_\mathrm{LMC}$ disk displays a clear S-shaped warp (first and second panels). The direct collision between the galaxies at the second $G_\mathrm{SMC}$ passage ($d_{\mathrm{LMC} - \mathrm{SMC}} = 0.26$~kpc) exerts an additional tidal torque (fourth and fifth panels) to the $G_\mathrm{LMC}$, heating and drastically distorting the whole disk.

\begin{figure}
  \resizebox{\hsize}{!}{\includegraphics{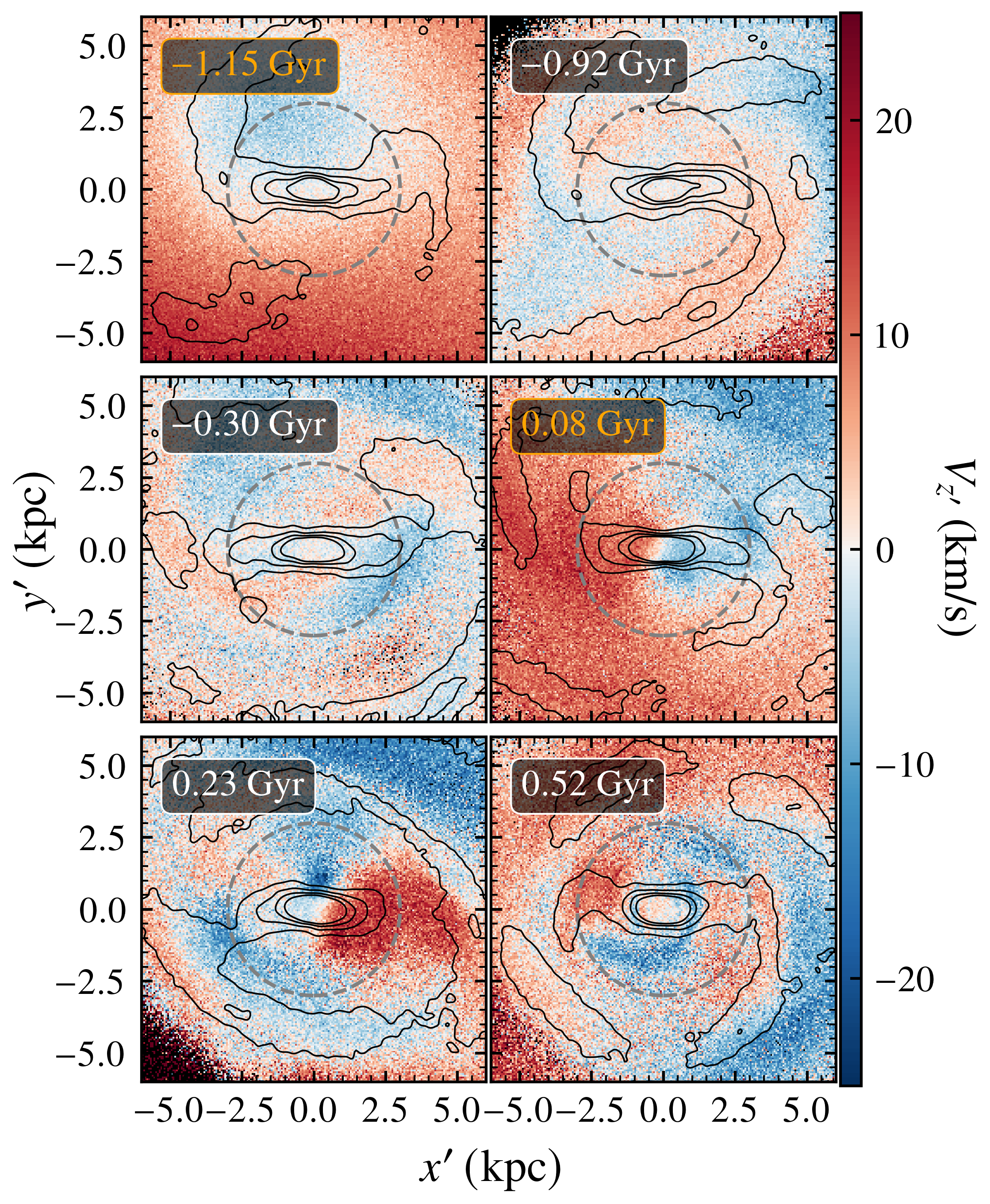}}
  \caption{Time evolution of the face-on mean vertical velocity map of the $G_\mathrm{LMC}$ in the K11 simulation. The text in the upper left corner denotes the time of the snapshot, while the orange text highlights the time at the first and second pericentric passages of $G_\mathrm{SMC}$. The $x^{\prime}$-axis is aligned with the semi-major axis of the bar. The gray circle denotes the region $R = 3$~kpc.  }
  \label{fig:K11_kinematics}
\end{figure}

To illustrate the kinematic evolution of the $G_\mathrm{LMC}$, we show the face-on mean vertical velocity map of the whole $G_\mathrm{LMC}$ in selected snapshots of the K11 simulation in \figref{fig:K11_kinematics}. Evidently, the two passages of the $G_\mathrm{SMC}$ (first and fourth panels) induce a dipole asymmetry in the $G_\mathrm{LMC}$ disk. After the first $G_\mathrm{SMC}$ passage, a weak quadrupole pattern emerges in the bar region of the $G_\mathrm{LMC}$ (third panel), characteristic of a buckling bar (fourth subplot of the left panels in \figref{fig:K11_bar_disk}, also see Fig.~7 in \citealt{2019A&A...629A..52L}). The direct collision with the $G_\mathrm{SMC}$ at the second passage (fourth panel) near the center of the $G_\mathrm{LMC}$ produces large vertical velocity asymmetries in both the $G_\mathrm{LMC}$ bar and disk. The quadrupole pattern in the bar region persists after the direct collision, though it becomes lopsided (fifth panel). The pattern winds up (sixth panel), as a boxy bulge forms at the center of the $G_\mathrm{LMC}$ (sixth subplot of the left panels in \figref{fig:K11_bar_disk}, also see \citealt{2019A&A...629A..52L}).

\subsection{The warped and buckling $G_\mathrm{LMC}$ bar}

\begin{figure}
  \resizebox{\hsize}{!}{\includegraphics{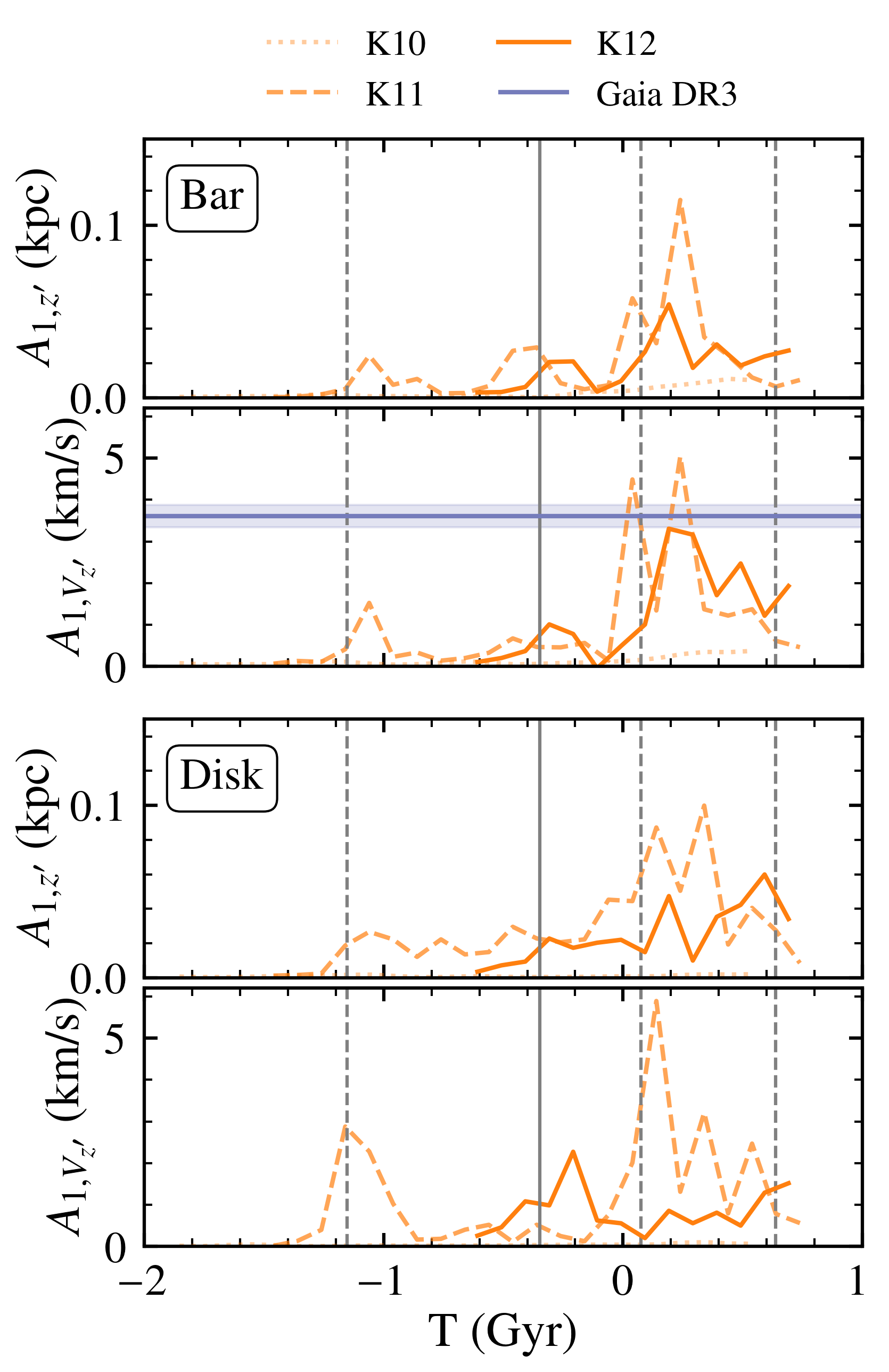}}
  \caption{From top to bottom: time evolution of the $m = 1$ vertical position (first and third panels) and velocity (second and fourth panels) asymmetry in the $G_\mathrm{LMC}$ bar (first and second panels) and disk (third and fourth panels) in the simulation set K10 (dotted), K11 (dashed) and K12 (solid lines). Uncertainty bars of the vertical asymmetries in the simulations obtained by bootstrapping $100$ times are approximately the size of the lines, so they are not shown in the figure. All lines are fitted using cubic splines as implemented in \code{scipy} \citep{scipy}. The $m = 1$ vertical velocity asymmetry measured in {\it Gaia}~DR3 observations is denoted by the horizontal purple line, where the shaded region denotes the uncertainties estimated by bootstrapping $100$ times. We discuss the vertical velocity asymmetries in {\it Gaia}~DR3 in Sect.~\ref{sec:Gaia}. The vertical grey (dashed) solid lines correspond to the pericenters of (K11) K12. }
  \label{fig:t_A1_zvz_bar_disk_smoothed}
\end{figure}

In order to quantify the detected vertical asymmetries in the $G_\mathrm{LMC}$ bar and disk in the previous subsection, we present the results of the Fourier decomposition method. We first study the vertical asymmetries in the $G_\mathrm{LMC}$ bar in the simulation set.

The top two panels of \figref{fig:t_A1_zvz_bar_disk_smoothed} shows the time evolution of the $m=1$ vertical position (first panel) and velocity (second panel) asymmetry in the $G_\mathrm{LMC}$ bar. Uncertainty bars of the vertical asymmetries in the simulations obtained by bootstrapping $100$ times are approximately the size of the lines, so they are not shown in the figure. The $G_\mathrm{LMC}$ bar in the isolated model, K10, shows $A_{1,z^{\prime}} \sim 0$ and $A_{1,V_{z^{\prime}}} \sim 0$ in the $G_\mathrm{LMC}$ bar throughout the simulation, indicative of vertical symmetry, as expected. Before the first $G_\mathrm{SMC}$ pericentric passage, the bars in interacting models K11 and K12 also have $A_{1,z^{\prime}} \sim 0$ and $A_{1,V_{z^{\prime}}} \sim 0$. For K11, both $G_\mathrm{SMC}$ passages result in a sudden increase in $A_{1,z^{\prime}}$ and $A_{1,V_{z^{\prime}}}$. The first peaks in $A_{1,z^{\prime}}$ and $A_{1,V_{z^{\prime}}}$ correspond to a slightly asymmetric $G_\mathrm{LMC}$ bar (first subplot of the left panels in \figref{fig:K11_bar_disk}). $A_{1,z^{\prime}}$ reaches another peak at $\sim-0.4$~Gyr, corresponding to the buckling $G_\mathrm{LMC}$ bar (third subplot of the left panels in \figref{fig:K11_bar_disk}). The second passage induces a significantly stronger $m = 1$ asymmetry due to the direct collision of the $G_\mathrm{LMC}$ with the SMC ($d_{\mathrm{LMC} - \mathrm{SMC}} = 0.26$~kpc), corresponding to a lopsided $G_\mathrm{LMC}$ bar (fourth subplot). The $m = 1$ vertical asymmetries in the bar significantly decrease after the collision, as a vertically symmetric bulge forms in the $G_\mathrm{LMC}$ (sixth subplot). Due to inclusion of the $G_\mathrm{MW}$ in K12, there is only one late pericentric passage of the $G_\mathrm{SMC}$. The single $G_\mathrm{SMC}$ passage in K12 occurs at a larger distance (see \figref{fig:orbit}, $d_{\mathrm{LMC} - \mathrm{SMC}} = 13.45$~kpc) compared to any $G_\mathrm{SMC}$ passages in K11. Therefore, the $m = 1$ vertical asymmetry in the $G_\mathrm{LMC}$ bar in K12 at the single $G_\mathrm{SMC}$ passage is smaller than those in K11.  
Approximately $400$~Myr after the $G_\mathrm{SMC}$ passage, both $A_{1,z^{\prime}}$ and $A_{1,V_{z^{\prime}}}$ in the bar in K12 reach a peak, corresponding to a warped bar.
The $m = 1$ vertical asymmetry in the $G_\mathrm{LMC}$ bar then oscillates but remains relatively large due to the buckling bar.

\begin{figure}
  \resizebox{\hsize}{!}{\includegraphics{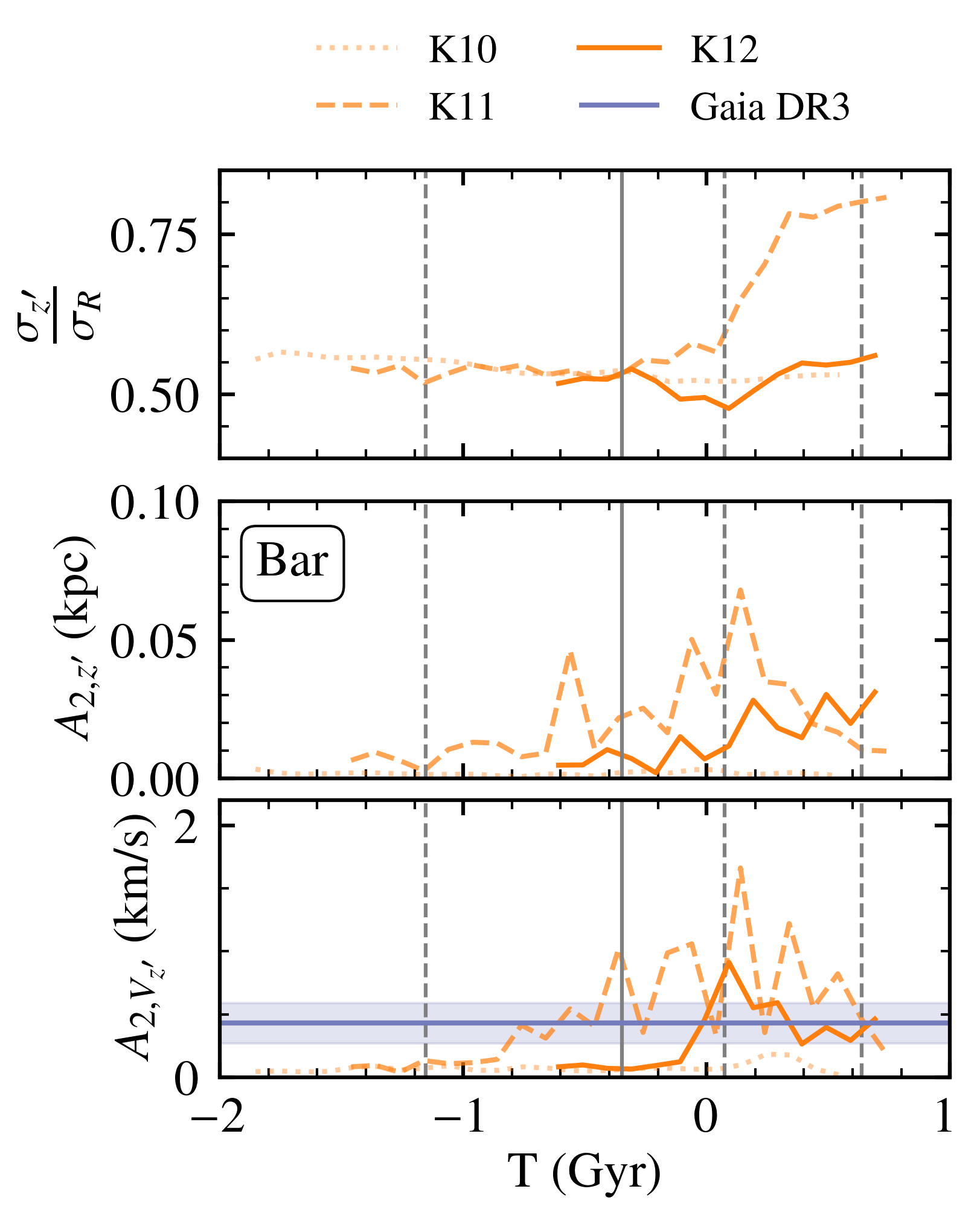}}
  \caption{Time evolution of the velocity dispersion ratio in the inner $G_\mathrm{LMC}$ ($R\leq 2$~kpc, top), as well as the $m = 2$ vertical position (middle) and velocity (bottom) asymmetry in the $G_\mathrm{LMC}$ bar in the simulation set K10, K11 and K12. Similar to \figref{fig:t_A1_zvz_bar_disk_smoothed}, uncertainties of the vertical asymmetries in the simulations obtained by bootstrapping are not shown in the figure. All lines are fitted using cubic splines. The $m = 2$ vertical velocity asymmetry measured in {\it Gaia}~DR3 observations is denoted by the horizontal purple line, and the discussion is deferred to Sect.~\ref{sec:Gaia}. }
  \label{fig:t_szsr_A2_zvz_bar}
\end{figure}

The top panel of \figref{fig:t_szsr_A2_zvz_bar} shows the time evolution of the velocity dispersion ratio $\sigma_z / \sigma_R$ in the inner $G_\mathrm{LMC}$ ($R\leq 2$~kpc). Buckling is characterized by an initial drop and a drastic rise of the velocity dispersion ratio \citep[e.g.,][]{2006ApJ...637..214M}. The velocity dispersion ratios of the inner $G_\mathrm{LMC}$ in both K11 and K12 display such characteristics. Combined with the morphological evolution (\figsref{fig:K11_bar_disk}), we conclude that the $G_\mathrm{LMC}$ bar in both K11 and K12 buckles after the first $G_\mathrm{SMC}$ passage (third subplot of the left panels in \figref{fig:K11_bar_disk}, 
while that in K10 never buckles. The middle and bottom panels of \figref{fig:t_szsr_A2_zvz_bar} track the time evolution of the $m=2$ vertical asymmetries in the $G_\mathrm{LMC}$ bar in the simulation set, typically associated with bar buckling \citep[e.g.,][]{2006ApJ...645..209D, 2019A&A...629A..52L}. For K11, the buckling bar corresponds to an increase in $A_{2,z^\prime}$ and $A_{2,V_{z^{\prime}}}$ after the first $G_\mathrm{SMC}$ passage. After the direct collision with the SMC, the $G_\mathrm{LMC}$ bar buckles violently, corresponding to a drastic increase of the velocity dispersion ratio and a peak in both $A_{2,z^{\prime}}$ and $A_{2,V_{z^{\prime}}}$. For K12, buckling happens approximately $400$~Myr after the first $G_\mathrm{SMC}$ passage, corresponding to a continuous rise in $A_{2,z^\prime}$ and a peak in $A_{2,V_{z^{\prime}}}$.

\subsection{The warped $G_\mathrm{LMC}$ disk}

The bottom two panels of \figref{fig:t_A1_zvz_bar_disk_smoothed} shows the time evolution of the $m=1$ vertical asymmetry in the $G_\mathrm{LMC}$ disk in the simulation set. The $G_\mathrm{LMC}$ disk in K10 remains flat throughout the simulation, as expected. Similarly to what happens to the $G_\mathrm{LMC}$ bar, each $G_\mathrm{SMC}$ passage results in a sudden rise in $A_{1, z^\prime}$ and $A_{1, V_{z^{\prime}}}$ for the interacting models K11 and K12. The increase in $A_{1, z^\prime}$ corresponds to a warped $G_\mathrm{LMC}$ disk (e.g., second and fourth subplots of the right panels in \figref{fig:K11_bar_disk}). For K11, the direct impact at the second $G_\mathrm{SMC}$ passage results in a larger warp in the $G_\mathrm{LMC}$ disk ($A_{1, z^\prime} \sim 0.1$~kpc, fourth subplot of the right panels in \figref{fig:K11_bar_disk}). For K12, $A_{1, V_{z^{\prime}}}$ gradually decreases as the $G_\mathrm{LMC}$ disk settles after the first $G_\mathrm{SMC}$ passage. $A_{1, z^\prime}$ and $A_{1, V_{z^{\prime}}}$ rise again near the end of the simulation, possibly due to a future second passage of the SMC, as suggested by \figref{fig:orbit}.

\figrefalt{fig:t_A1_zvz_bar_disk_smoothed} provides a comparison between the $m=1$ vertical asymmetry in the $G_\mathrm{LMC}$ bar and that in the disk. The plot demonstrates that the $G_\mathrm{SMC}$ passage induces $m = 1$ vertical asymmetries in both the $G_\mathrm{LMC}$ bar and disk. For both interacting models, $A_{1, V_{z^{\prime}}}$ in the disk is larger than that in the bar at the first $G_\mathrm{SMC}$ passage. For K11, $A_{1, V_{z^{\prime}}}$ in the bar increases and becomes similar to that in the disk at the direct impact of the $G_\mathrm{SMC}$ ($d_{\mathrm{LMC} - \mathrm{SMC}} = 0.26$~kpc). For K12, $A_{1,V_{z^{\prime}}}$ in the $G_\mathrm{LMC}$ disk gradually returns to $\sim 0$ after the first $G_\mathrm{SMC}$ passage as the disk settles, while that in the $G_\mathrm{LMC}$ bar continues to rise, due to bar buckling. Cross-correlation analysis reveals that the increase in $A_{1,V_{z^{\prime}}}$ in the $G_\mathrm{LMC}$ bar and disk at each $G_\mathrm{SMC}$ pericentric passage occurs without significant delay --- the $G_\mathrm{LMC}$ bar and disk respond to each $G_\mathrm{SMC}$ passage simultaneously.

\section{The $m = 1$ asymmetry in the $G_\mathrm{LMC}$ bar across the KRATOS suite}
\label{sec:all_models}
\begin{figure}
  \resizebox{\hsize}{!}{\includegraphics{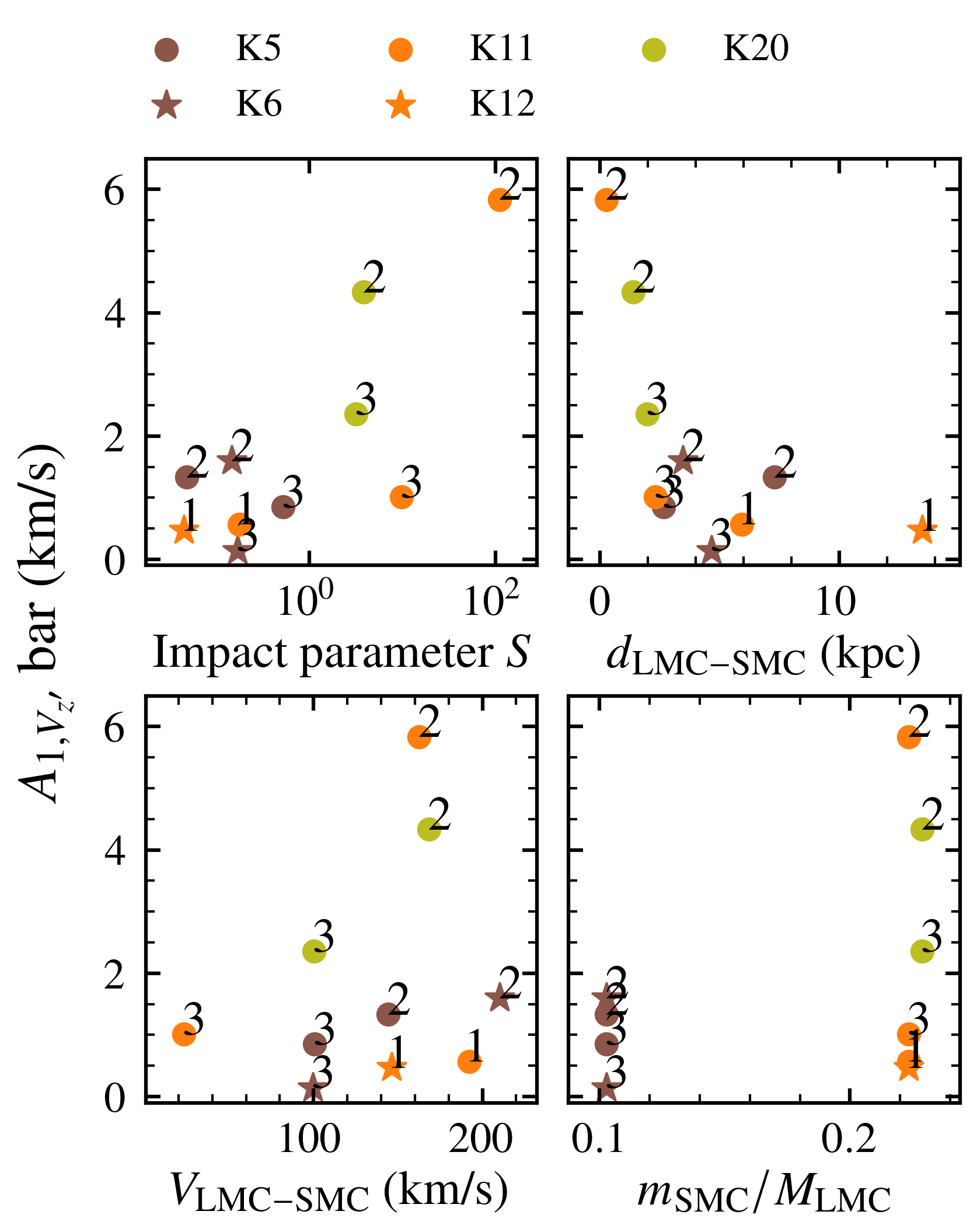}}
  \caption{Amplitude of the $m = 1$ $V_{z^{\prime}}$ asymmetry versus the impact parameter $S$ (in log scale), distance between $G_\mathrm{LMC}$ and SMC, their relative velocity, and their mass ratio at each $G_\mathrm{SMC}$ passage, for all interacting simulation models. Circles denote the $G_\mathrm{LMC}+G_\mathrm{SMC}$ models, while stars denote the $G_\mathrm{LMC}+G_\mathrm{SMC} + G_\mathrm{MW}$ models. The number label denotes the $i$-th passage of the $G_\mathrm{SMC}$. The K21 simulation has not formed a bar at the first $G_\mathrm{SMC}$ passage, so it is not considered in this plot. }
  \label{fig:passage_A1_vz_bar}
\end{figure}

To understand the cause of $m = 1$ vertical velocity asymmetry in the $G_\mathrm{LMC}$ bar (see \figref{fig:t_A1_zvz_bar_disk_smoothed}), we present a summary plot of $A_{1,V_{z^{\prime}}}$ in the bar at each $G_\mathrm{SMC}$ passage with respect to various parameters of tidal interactions and initial conditions for five interacting models in \figref{fig:passage_A1_vz_bar}. The dimensionless impact parameter $S$ \citep{1991A&A...244...52E} is calculated by
\begin{align}
    S = \frac{m_\mathrm{SMC}}{M_\mathrm{LMC}} \left( \frac{R_\mathrm{LMC}}{d_{\mathrm{LMC} - \mathrm{SMC}}} \right)^3 \frac{\Delta T}{P} \,,
\end{align}
where $m_\mathrm{SMC}/ M_{\mathrm{LMC}}$ is calculated by the ratio between the total mass (stellar component and dark matter) of the two galaxies, ${R_\mathrm{LMC}}$ is the disk truncation radius of the $G_\mathrm{LMC}$, $\Delta T$ is the interaction time, and $P$ is the time for the star particles in the outer $G_\mathrm{LMC}$ disk to rotate one radian in their orbits. $V_{\mathrm{LMC} - \mathrm{SMC}}$ is defined as the relative bulk velocity of the stellar component of the $G_\mathrm{LMC}$ and $G_\mathrm{SMC}$. The K21 simulation is neglected in the plot, because it has not formed a bar at the first $G_\mathrm{SMC}$ passage.

In the top left panel of \figref{fig:passage_A1_vz_bar}, the amplitude of the $m = 1$ vertical velocity asymmetries in the $G_\mathrm{LMC}$ bar at each $G_\mathrm{SMC}$ pericentric passage shows a clear correlation with the impact parameter $S$ (Spearman coefficient $r_s = 0.59$), indicating that the vertical velocity asymmetries are indeed induced by the $G_\mathrm{SMC}$ passages.

To assess the effect of each interaction parameter, we consider the other three panels. The top right panel shows that $A_{1,V_{z^{\prime}}}$ decreases with the distance between the $G_\mathrm{LMC}$ and the $G_\mathrm{SMC}$, with a Spearman coefficient of $r_s = -0.77$. As an example, at the second $G_\mathrm{SMC}$ passage, $A_{1,V_{z^{\prime}}}$ in the bar for the $G_\mathrm{LMC}+G_\mathrm{SMC}$ models K5, K11 and K20 show a clear declining trend with $d_{\mathrm{LMC} - \mathrm{SMC}}$. At the third passage, $A_{1,V_{z^{\prime}}}$ in the bar in K20 is larger than that in K5 or K11, though the distance between the two galaxies is similar. This is likely due to the higher relative velocity of the galaxies at the third passage in K20 than in K11, and the larger mass ratio in K20 than in K5 or K11. For K20, $A_{1,V_{z^{\prime}}}$ at the third $G_\mathrm{SMC}$ passage is smaller than that at the second passage, though the two passages occur at similar distances. This is likely due to the lower relative velocity at the third passage. For the $G_\mathrm{LMC}+G_\mathrm{SMC}+G_\mathrm{MW}$ models K6 and K12, $A_{1,V_{z^{\prime}}}$ in the bar is generally smaller, either due to the large distance between the galaxies (K12), or the low mass ratio of the galaxies (K6).

In the bottom panels of \figref{fig:passage_A1_vz_bar}, $A_{1,V_{z^{\prime}}}$ in the bar does not correlate strongly with the relative velocity of the galaxies ($r_s = 0.31$), or the mass ratio between the two galaxies ($r_s = 0.44$). Note that the galaxies at all passages in the same simulation set have the same initial mass ratio. However, we observe that $G_\mathrm{SMC}$ passages in simulations with the lowest mass ratio (K5 and K6) generally induce small $A_{1,V_{z^{\prime}}}$ in the bar.

\figrefalt{fig:passage_A1_vz_bar} suggests that $d_{\mathrm{LMC} - \mathrm{SMC}}$ is the dominant interaction parameter for determining the amplitudes of the vertical asymmetries in the $G_\mathrm{LMC}$ bar. $V_{\mathrm{LMC} - \mathrm{SMC}}$ and $m_\mathrm{SMC}/ M_{\mathrm{LMC}}$ only weakly affect the amplitudes of the induced vertical asymmetries.

\section{Discussion}
\label{sec:discuss}
In this section, we compare our results with the simulations in \citet[][Sect.~\ref{sec:simu}]{2012MNRAS.421.2109B} and put our results in context with respect to {\it Gaia}~DR3 observations (Sect.~\ref{sec:Gaia}). We estimate possible biases and systematics by applying the same method to {\it Gaia} mock catalogs (Sect.~\ref{sec:mocks}).

\subsection{Qualitative comparison with \citet{2012MNRAS.421.2109B}}
\label{sec:simu}
Here, we compare the results in this work with the ones from Model 2 in \citet{2012MNRAS.421.2109B}, and the more recent work by \citet{2025ApJ...988...79R} using the same simulation. While they perform a similar analysis to ours, there are some differences between their simulations and models and KRATOS, which are worth mentioning before the direct comparison.

First, at present time, Model 2 of \citet{2012MNRAS.421.2109B} consists of a tilted bar and a warped disk. Second, the authors consider the bar as a rigid body, in particular, \citet{2025ApJ...988...79R} measure the angle between the ellipse fitted to the rigid bar and the disk plane in a side-on projection. It is also worth mentioning that the authors do not compare with simulations of an isolated $G_\mathrm{LMC}$, or a $G_\mathrm{LMC}$ interacting with only a $G_\mathrm{SMC}$; or with simulations of varying initial conditions. Finally, \citet{2012MNRAS.421.2109B} used an $N$-body SPH simulation, while we use representative pure $N$-body simulations from KRATOS. Because the LMC is a gas-rich galaxy, gas dynamics could play an important role in its evolution. A future $N$-body SPH version of the KRATOS simulations will shed light on the effect of gas dynamics (Jim\'enez-Arranz, Roca-F\`abrega et al., in prep.).

In this study, we track the morphological evolution of the $G_\mathrm{LMC}$ bar after extracting the bar particles. We find that the $G_\mathrm{SMC}$ pericentric passages induce vertical asymmetries in the $G_\mathrm{LMC}$ bar, but have not observed a rigid tilt in the $G_\mathrm{LMC}$ bar with respect to the $G_\mathrm{LMC}$ disk plane for all seven simulation models. In particular, the $G_\mathrm{LMC}$ bar may be deformed by buckling, and the $G_\mathrm{LMC}$ disk may also be severely disrupted. For example, the fifth subplot of \figref{fig:K11_bar_disk} ($\sim 150$~Myr after the direct collision of the galaxies) is qualitatively similar to the middle right panel in Fig.~13 in \citet{2012MNRAS.421.2109B} or the right panel in Fig.~2 in \citet{2025ApJ...988...79R}. Our analysis reveals that the direct collision with the $G_\mathrm{SMC}$ induces an $m = 1$ vertical asymmetry to the already buckling $G_\mathrm{LMC}$ bar.

\begin{figure}
\centering
  \resizebox{\hsize}{!}{\includegraphics{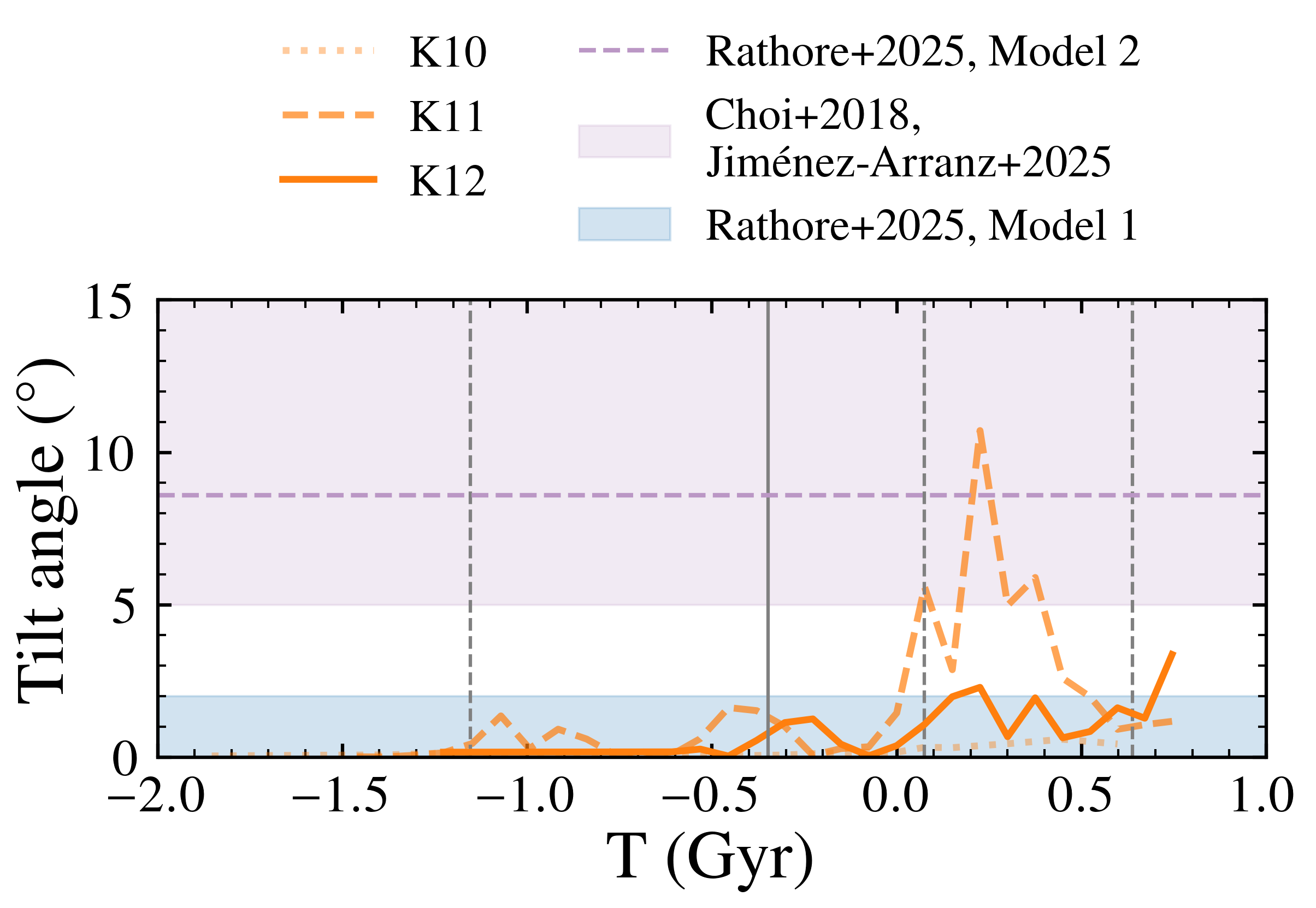}}
  \caption{Time evolution of the tilt of the $G_\mathrm{LMC}$ bar in the simulation set K10 (dotted), K11 (dashed) and K12 (solid line). The tilt angle is estimated by the angle between the principal direction of the $G_\mathrm{LMC}$ bar and the disk plane ($z^{\prime} = 0$). The purple shaded region shows the estimated tilt of the $G_\mathrm{LMC}$ bar in \citet{2018ApJ...866...90C} and \citet{2025A&A...698A..88J}. The purple dashed line denotes the tilt angle of the $G_\mathrm{LMC}$ bar at the present time ($\sim 8.6\degree$) in Model 2 of \citet{2025ApJ...988...79R}. The blue shaded region shows the $1\sigma$ spread around the mean tilt angle of the $G_\mathrm{LMC}$ bar ($\sim 1\degree$) in Model 1 of \citet{2025ApJ...988...79R}. }
  \label{fig:t_tilt_K11}
\end{figure}

Because the $G_\mathrm{LMC}$ bar and disk are both deformed by tidal interactions, we cannot obtain a rigid tilt angle. We can, however, estimate a ``tilt angle" by measuring the angle between the principal direction of the bar obtained by principal component analysis (PCA) and the disk plane ($z^{\prime} = 0$). \figrefalt{fig:t_tilt_K11} illustrates the time evolution of the estimated tilt angle of the $G_\mathrm{LMC}$ bar in the simulation set K10, K11 and K12, which can be compared to Fig.~8 in \citet{2025ApJ...988...79R}. The tilt angle of the bar at the present time ($\sim 8.6\degree$) in Model 2 of \citet{2025ApJ...988...79R} is denoted by the purple dashed line, while the $1\sigma$ spread around the mean tilt angle of the bar in Model 1 ($\sim 1\degree$) is illustrated by the blue shaded region. In \figrefalt{fig:t_tilt_K11}, the tilt of the $G_\mathrm{LMC}$ bar in the isolated model K10 remains below $\sim 1\degree$ throughout the simulation, consistent with Model~1. The $G_\mathrm{LMC}$ bar in K12 does not show a significant tilt, either, being consistently within the $1\sigma$ spread around the mean tilt angle of the bar in Model~1 before the end of the simulation. In K11, the first and third $G_\mathrm{SMC}$ passages do not induce a significant tilt to the $G_\mathrm{LMC}$ bar, while the direct collision with the $G_\mathrm{SMC}$ at the second passage induces a tilt within the range of $5 \text{-} 15 \degree$, consistent with observations \citep[e.g.,][]{2018ApJ...866...90C,2025A&A...698A..88J}. The maximum tilt of the $G_\mathrm{LMC}$ bar ($10.7\degree$) occurs $\sim 150$~Myr after the direct collision, slightly larger than that in \citet[][$\sim 9.5\degree$]{2025ApJ...988...79R}, possibly due to the smaller distance between the galaxies at the direct collision in this work ($d_{\mathrm{LMC} - \mathrm{SMC}} = 0.26$~kpc) compared to that in \citet[][$d_{\mathrm{LMC} - \mathrm{SMC}} \sim 2$~kpc]{2025ApJ...988...79R}. The tilt angle remains observationally significant for $\sim 300$~Myr after the direct collision, consistent with the current estimate of the recent close encounter between the LMC and the SMC \citep[$150-300$\,Myr ago, e.g.,][]{2011ApJ...737...29O,2022ApJ...927..153C, 2022MNRAS.510..445C}.

The estimated tilt of the $G_\mathrm{LMC}$ bar in the other four simulation models is not observationally significant (see \figref{fig:t_tilt_K4_K19} in App.~\ref{sec:supplementary}). In particular, the tilt in the models with low mass ratio between the $G_\mathrm{LMC}$ and the $G_\mathrm{SMC}$ ($m_\mathrm{SMC} /{M_\mathrm{LMC}} \sim 0.10$, compared to $m_\mathrm{SMC} /{M_\mathrm{LMC}} \sim 0.22$ for the other models), K5 and K6, does not exceed the $1\sigma$ spread around the mean tilt angle of the bar in Model~1 throughout the simulation. This is consistent with \figref{fig:passage_A1_vz_bar} and the discussion in Sect.~\ref{sec:all_models}. Although we are unable to explore the full range of masses of the $G_\mathrm{LMC}$ and the $G_\mathrm{SMC}$ due to the limitation of the KRATOS simulations, the current result suggests that models with a higher mass ratio between the galaxies may be favorable for reproducing the observed tilt of the LMC bar.

\subsection{Comparison with {\it Gaia}~DR3 observation}
\label{sec:Gaia}
\begin{figure}
  \resizebox{\hsize}{!}{\includegraphics{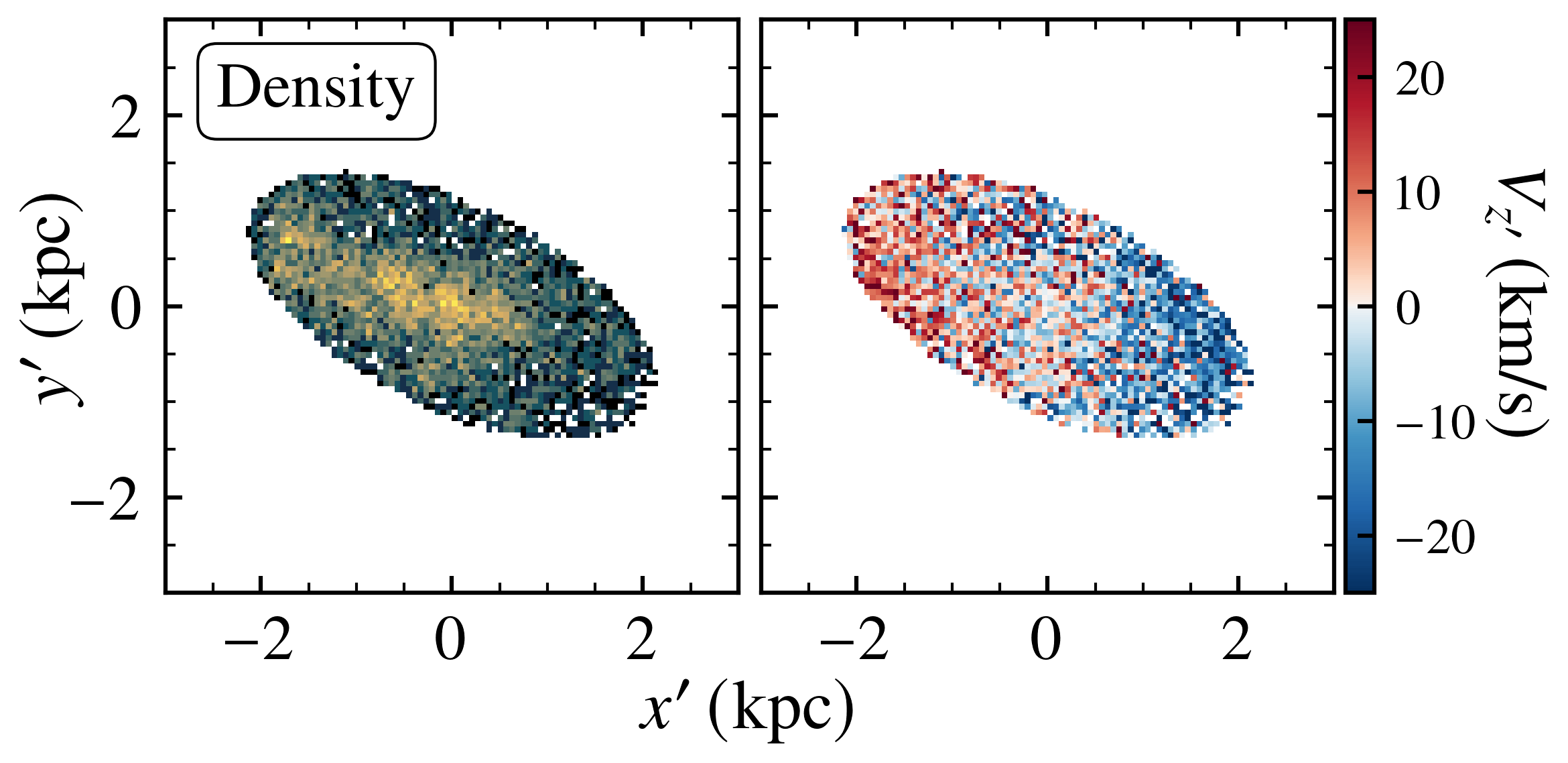}}
  \caption{The face-on view of the LMC bar in the \textit{Gaia}~DR3 observation color-coded by density (left) and vertical velocity (right). The LMC bar is obtained using the ellipse fitting method in \citet{2025A&A...701A.227S}.}
  \label{fig:gaia_bar}
\end{figure}

Because the {\it Gaia}~DR3 observations lack accurate distance measurements for the LMC stars, we hereafter focus on the analysis of the vertical velocity asymmetries in the LMC.
We use the sample of LMC stars with available line-of-sight velocities in {\it Gaia}~DR3 from \citet{2023A&A...669A..91J} for this analysis. The observed LMC stars are projected onto the LMC disk plane ($z^\prime = 0$) according to \citet{2023A&A...669A..91J}. The limited size of the sample ($N = \numprint{30749}$) makes it difficult to apply our topology-based bar classification method. Instead, following \citet{2025A&A...701A.227S}, we use the ellipse fitting method to obtain the observed LMC bar, which contains $N = \numprint{9959}$ stars. The face-on density (left) and vertical velocity maps (right) of the obtained LMC bar are shown in \figref{fig:gaia_bar}. In the right panel of \figref{fig:gaia_bar}, the vertical velocities of the LMC bar stars in the {\it Gaia}~DR3 sample display a clear dipole asymmetry.

We perform Fourier decomposition and illustrate the {\it Gaia}~DR3 results as horizontal purple lines in \figsref{fig:t_A1_zvz_bar_disk_smoothed} and \ref{fig:t_szsr_A2_zvz_bar}. Uncertainties are estimated by bootstrapping \numprint{100} times, and therefore primarily reflects statistical errors. As a result, the total error is likely underestimated. \figrefalt{fig:t_A1_zvz_bar_disk_smoothed} shows that compared with the simulation models K11 and K12, there exists significant $m = 1$ vertical velocity asymmetry in the observed LMC bar in {\it Gaia}~DR3. In particular, the $m = 1$ vertical velocity asymmetry in the $G_\mathrm{LMC}$ bar in K11 remains consistent with the {\it Gaia}~DR3 observations (within $1\sigma$ of the observed value) $\sim 210$~Myr after the direct collision. The observed $m = 1$ vertical asymmetries are therefore in agreement with the recent passage of the SMC \citep[e.g.,][]{2011ApJ...737...29O,2022ApJ...927..153C,2022MNRAS.510..445C}. The significant $m = 1$ vertical asymmetries in the observed LMC bar are further supported by the {\it Gaia} mock observations in the following subsection.

In \figref{fig:t_szsr_A2_zvz_bar}, $A_{2,V_{z^{\prime}}}$ in the observation is relatively low compared to the simulations. Therefore, there is no clear $m = 2$ vertical asymmetry in the observed LMC bar.

The face-on density, radial, azimuthal, and vertical velocity maps of the LMC sample can be referred to in Figs.~15, 16 and 17 of \citet{2023A&A...669A..91J}. For the LMC disk ($3\leq R \leq 6$~kpc), we obtain $A_{1,V_{z^{\prime}}} = 1.92 \pm 0.38 \kms$ and $A_{2,V_{z^{\prime}}} = 3.29 \pm 0.32 \kms$. The observed disk asymmetries are significant compared to the simulations (see \figsref{fig:t_A1_zvz_bar_disk_smoothed} and~\ref{fig:t_szsr_A2_zvz_bar}).

\subsection{{\it Gaia} mock catalogs}
\label{sec:mocks}
In this section, we present the {\it Gaia} mock catalogs constructed from the simulations. We use {\it Gaia} mock catalogs to illustrate the limitations of our method and to provide caveats for interpreting the values we obtained from {\it Gaia}~DR3 observations reported in Sect.~\ref{sec:Gaia}.

We simulate {\it Gaia} observables and uncertainties for the KRATOS simulations using the mock catalog tool presented in \citet{2025A&A...701A.227S} \footnote{https://github.com/mschoelch24/MockCatalogue}.
We start by projecting the KRATOS simulations (centered and aligned as stated in Sect.~\ref{sec:cen_align}) to the position and orientation of the LMC in the heliocentric reference frame. We use center coordinates ($\alpha_0, \delta_0$) = ($81.28^\circ, -69.78^\circ$) according to \citet{vandermarel01}, and orient the simulation according to the LMC inclination angle $i = 34^\circ$ and position angle $\theta = 220^\circ$ \citep{GaiaLuri21}. Then we apply a 3D galactic extinction model by \citet{marshall06} and \citet{lallement22}. Since the LMC stars are located beyond the model's distance limit of 15 kpc, the applied extinctions correspond to the foreground extinction of the MW. 
To simulate observables of the stars in the LMC, we use a high purity LMC sample from \citep[][]{2023A&A...669A..91J} that has line-of-sight velocities available, and model the distribution of apparent color $G_{BP} - G_{RP}$, absolute magnitude $M_G$, and radial velocity uncertainty $\sigma_{V_{los}}$ with a kernel density estimation (KDE). We then draw randomly from this KDE to assign the observational parameters to the simulation particles, and from the absolute magnitude, color, and extinction we derive apparent $G$ magnitudes. Then we use the {\it Gaia}~DR3 error model also implemented in \textsc{PyGaia} \citep{pygaia,gaiadr3summary} to compute astrometric uncertainties in position ($\sigma_{\alpha*}, \sigma_{\delta}$), parallax ($\sigma_{\varpi}$) and proper motion ($\sigma_{\mu_{\alpha*}}, \sigma_{\mu_{\delta}}$). We apply the line-of-sight velocity uncertainties ($\sigma_{V_{los}}$) corresponding to the color and absolute magnitude from the KDE described above to model the uncertainties directly on the LMC line-of-sight velocity distribution. 
In {\it Gaia} DR3, line-of-sight velocities are available only for stars with $G_{RVS} \leq 14$  \citep{Katz23}. To apply a comparable cutoff to our simulated stars, we compute the $G_{RVS}$ magnitudes using the color-color relation between $G - G_{RVS}$ and $G_{BP} - G_{RP}$ from \citet{Sartoretti22} and exclude all stars above $G_{RVS} = 14$ from the mock catalog. 
Using the derived uncertainties for each star particle, we sample a new value for the position ($\alpha, \delta$), parallax ($\varpi$), proper motions ($\mu_{\alpha*}, \mu_{\delta}$), and line-of-sight velocities ($V_{los}$) from a Gaussian distribution centered on the true simulated value, with standard deviation of the computed uncertainty. Lastly, we convert the {\it Gaia} mock catalog back to the deprojected $G_\mathrm{LMC}$ in-plane coordinate frame \citep[for details, see Appendix A of][]{2023A&A...669A..91J}. Due to the large uncertainties in the parallaxes at the distance of the Clouds, we assume all stars are located in an infinitely thin disk and have $z' = 0$ in the LMC in-plane coordinates.

\begin{figure}
\resizebox{\hsize}{!}{\includegraphics{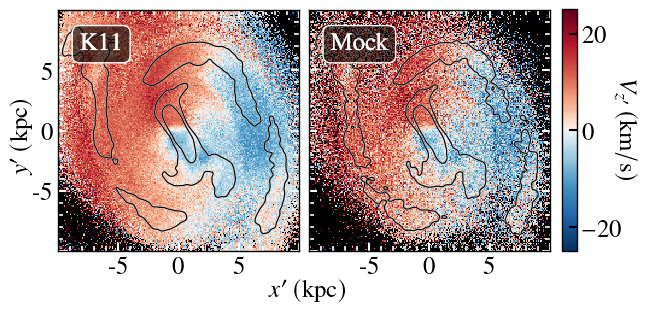}}
  \caption{Vertical velocity maps of the K11 simulation at $t = 0.075$~Gyr (left panel), and its corresponding mock catalog (right panel).}
  \label{fig:K11mock}
\end{figure}

In Fig.~\ref{fig:K11mock}, we show a comparison of the vertical velocity map of the K11 simulation at $t = 0.075$~Gyr (left panel) with its corresponding mock catalog (right panel), predicting the vertical signatures found in {\it Gaia} data. The snapshot was previously aligned according to the method described in Sect. \ref{sec:cen_align} and presents the strongest asymmetry signatures due to the second $G_\mathrm{SMC}$ passage (\figref{fig:passage_A1_vz_bar}). 
The asymmetric signature in the vertical velocity map is clearly present in this simulation snapshot, and the mock catalog reproduces velocities across the full disk.

Applying the same Fourier decomposition, we obtain $A_{1,V_{z^{\prime}}} = 5.71 \pm 0.13 \kms$ for the bar, and $A_{1,V_{z^{\prime}}} = 4.09 \pm 0.04 \kms$ for the disk region of the {\it Gaia} mock catalog, where the uncertainty is estimated by bootstrapping and therefore primarily reflects statistical errors. As a result, the total error is likely underestimated and the results are reasonably consistent with the simulation ($A_{1,V_{z^{\prime}}} = 5.84 \pm 0.07 \kms$ for the bar, $A_{1,V_{z^{\prime}}} = 4.27 \pm 0.02 \kms$ for the disk region).

From the mock catalogs, we can conclude that {\it Gaia} astrometric uncertainties and the lack of individual heliocentric distances for each source do not bias or create artificial systematics in the velocity maps. The main effect of these limitations is some blurring of the kinematic signatures in the outskirts of the disk, but the main trends and kinematic asymmetries remain visible.

\section{Conclusions}
\label{sec:conclude}
In this work, we presented the vertical asymmetries detected in the $G_\mathrm{LMC}$ in seven $N$-body simulations of the KRATOS suite. We tracked the morphological and kinematic evolution of the $G_\mathrm{LMC}$ bar and disk, and provided quantitative analysis of the detected asymmetries. The main results of this work are:
\begin{itemize}
\item
We tracked the morphological evolution of the $G_\mathrm{LMC}$ bar and disk separately (\figref{fig:bar_classification}) using a topology-based classification method. We concluded that tidal interactions between the two interacting galaxies result in a warped and buckling $G_\mathrm{LMC}$ bar and a warped $G_\mathrm{LMC}$ disk (\figref{fig:K11_bar_disk}). Tidal interactions also induce vertical asymmetries in the kinematics of the $G_\mathrm{LMC}$ (\figref{fig:K11_kinematics}).

\item 
We quantified the detected vertical asymmetries in the bar and disk of the $G_\mathrm{LMC}$ using Fourier decomposition (\figsref{fig:t_A1_zvz_bar_disk_smoothed} and \ref{fig:t_szsr_A2_zvz_bar}), and found that the induced vertical asymmetries in the bar and disk of the $G_\mathrm{LMC}$ occur simultaneously.

\item 
We performed correlation analysis with respect to various interaction parameters and initial conditions, which revealed that the impact parameter $S$ predicts well the amplitude of the vertical asymmetries, while the distance between the galaxies has the strongest effect on the induced asymmetries (\figref{fig:passage_A1_vz_bar}).

\item 
Compared to Model 1 and Model 2 of the $N$-body SPH simulation in \citet{2012MNRAS.421.2109B}, we did not detect a rigidly tilted bar in all seven simulation models, but detected a $G_\mathrm{LMC}$ bar deformed by the pericentric passages of the $G_\mathrm{SMC}$ through buckling and warping. The differences between our results may be attributed to the gas dynamics, considering that the LMC is a gas-rich galaxy. A future $N$-body SPH version of the KRATOS simulations (Jim\'enez-Arranz, Roca-F\`abrega et al., in prep.) will shed light on the effect of gas dynamics.

\item
By estimating a tilt angle of the $G_\mathrm{LMC}$ bar using PCA, we discovered that the direct collision between the galaxies in our high mass ratio model ($m_\mathrm{SMC} /{M_\mathrm{LMC}} \sim 0.22$), K11, produces an observationally significant tilt ($10.7\degree$, \figref{fig:t_tilt_K11}). 
Tidal interactions in our lower mass ratio models ($m_\mathrm{SMC} /{M_\mathrm{LMC}} \sim 0.10$), K5 and K6, however, fail to produce a significant tilt of the $G_\mathrm{LMC}$ bar throughout the simulation (\figref{fig:t_tilt_K4_K19}). Thus, models with a higher mass ratio between the SMC and the LMC may be favorable for reproducing the observed tilt of the LMC bar.

\item
In K11, after the direct collision between the $G_\mathrm{LMC}$ and the $G_\mathrm{SMC}$, the estimated tilt angle remains observationally significant (within the range of $5\degree \text{-} 15\degree$) for $\sim 300$~Myr.

\item 
We observed significant $m = 1$ asymmetry in the vertical velocity map of the LMC bar in {\it Gaia}~DR3 (\figref{fig:gaia_bar}) compared to the simulations (\figref{fig:t_A1_zvz_bar_disk_smoothed}). In particular, the observed $m = 1$ vertical asymmetry remains significant compared with the K11 simulation $\sim 210$~Myr after the direct collision. Therefore, our estimated upper limit for the time since the most recent SMC pericentric passage is approximately $210$ to $300$~Myr, consistent with the current observational estimate of the recent passage of the SMC (approximately $150-300$\,Myr ago). 

\item
Our {\it Gaia}~DR3 mock catalogs are able to reproduce asymmetric kinematic signals from the simulations (\figref{fig:K11mock}) and therefore suggest that astrometric uncertainties in the {\it Gaia} data do not bias the results. 

\end{itemize}

In summary, we conclude that tidal interactions between the $G_\mathrm{LMC}$ and the $G_\mathrm{SMC}$ induce significant vertical asymmetries in the $G_\mathrm{LMC}$ bar and disk in the KRATOS simulations. We detect significant vertical asymmetries in the LMC bar in the {\it Gaia}~DR3 observations, which is in agreement with the recent passage of the SMC. The upcoming {\it Gaia}~DR4\footnote{https://www.cosmos.esa.int/web/gaia/dr4} and the $N$-body SPH version of the KRATOS simulations will allow us to gain better understanding of the vertical asymmetries in the bar and disk of the LMC.

\begin{acknowledgements}
We thank the anonymous referee for providing suggestions that improved this paper. SW greatly appreciates the helpful discussions at the ``Galactic bars, bulges and disks" workshop in Lipari. SW is grateful to P. S\'anchez-Mart\'in for her help with the bar classification algorithm.

This project has received funding from the European Union's Horizon 2020 research and innovation programme under the Marie Sk{\l}odowska-Curie grant agreement No. 101072454 (MWGaiaDN).

OJA acknowledges funding from ``Swedish National Space Agency 2023-00154 David Hobbs The GaiaNIR Mission'' and ``Swedish National Space Agency 2023-00137 David Hobbs The Extended Gaia Mission.''

This work has made use of data from the European Space Agency (ESA) mission {\it Gaia} (\url{https://www.cosmos.esa.int/gaia}), processed by the {\it Gaia} Data Processing and Analysis Consortium (DPAC, \url{https://www.cosmos.esa.int/web/gaia/dpac/consortium}). Funding for the DPAC has been provided by national institutions, in particular the institutions participating in the {\it Gaia} Multilateral Agreement.

This work made use of the following software: \code{scikit-learn} \citep{scikit-learn}, \code{GUDHI} \citep{gudhi:AlphaComplex}, \code{NetworkX} \citep{SciPyProceedings_11}, \code{astropy} \citep{astropy:2013,astropy:2018}, \code{matplotlib} \citep{matplotlib}, \code{scipy} \citep{scipy}, \code{numpy} \citep{numpy}.

\end{acknowledgements}

\bibliographystyle{aa}
\bibliography{main}{}

\begin{appendix}

\section{Bar classification}
\subsection{The topology-based bar classification method}
\label{sec:bar_class}
In this section, we present the details and software used when applying the topology-based bar classification method in Sect.~\ref{sec:method}. As an example, we show the steps of applying the method to the K11 simulation at $t = 0.075$~Gyr (the second $G_\mathrm{SMC}$ pericentric passage).

\begin{figure}
  \resizebox{\hsize}{!}{\includegraphics{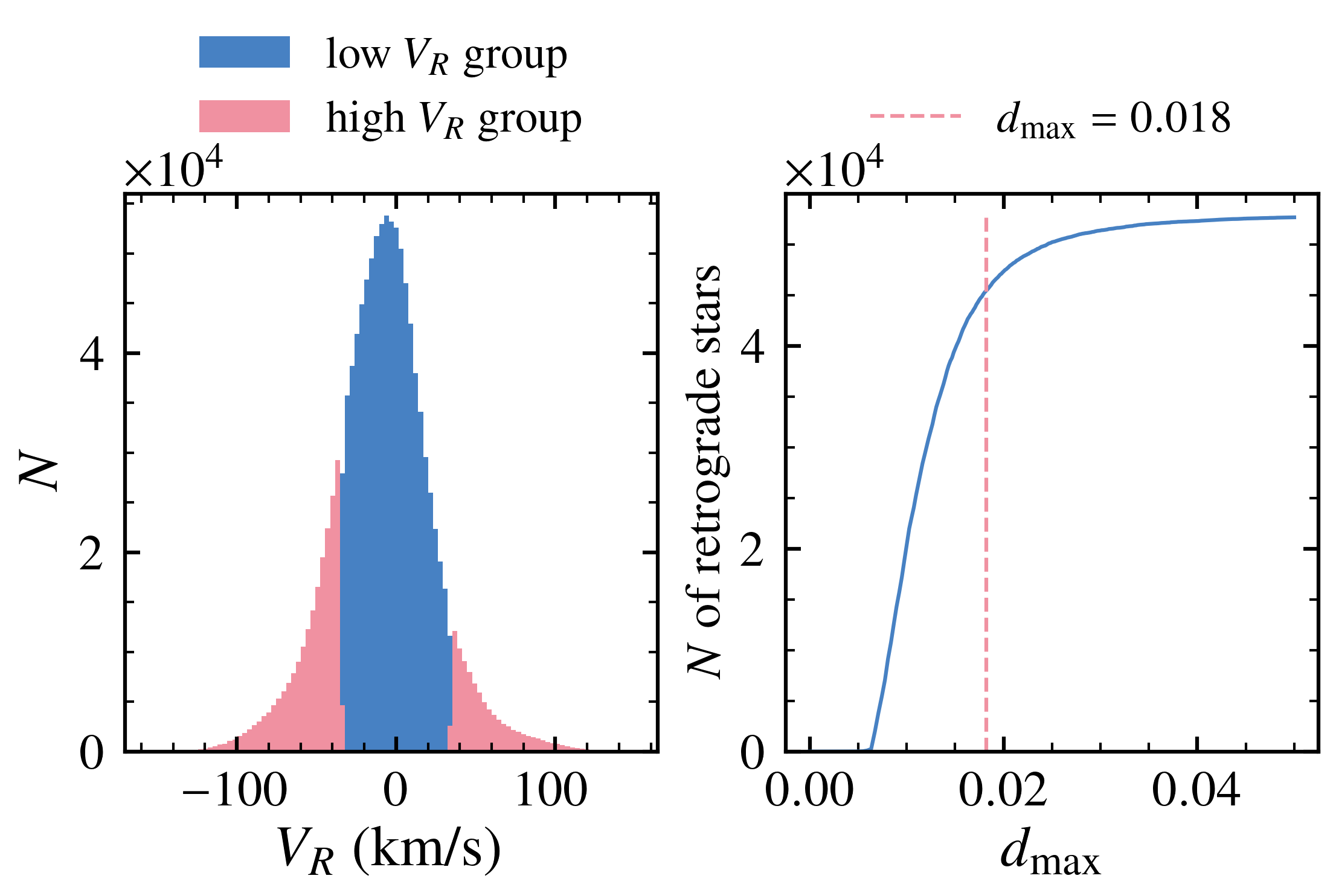}}
  \caption{Left: histogram of the radial velocities of the $G_\mathrm{LMC}$ star particles in the K11 simulation at $t = 0.075$~Gyr, after applying the K-means clustering algorithm. Pink highlights the high radial velocity group. Right: $d_\mathrm{max}$ versus the number of retrograde star particles within the LCC. The pink dashed line corresponds to the knee of the plot.}
  \label{fig:kmeans_dmax_retro}
\end{figure}

The left panel in \figref{fig:kmeans_dmax_retro} shows the histogram of the radial velocities of the $G_\mathrm{LMC}$ star particles. We use \code{scikit-learn} to perform K-means clustering, dividing the $G_\mathrm{LMC}$ star particles into a high radial velocity (pink) and a low radial velocity (blue) group. Three dimensional positions of the star particles belonging to the high radial velocity group form the point cloud in our topological analysis.

Persistent homology studies how topological features such as connected components evolve as the topological space changes. For simplicity, we consider the alpha complex version of persistent homology. We use \code{GUDHI} to construct the alpha complexes. We use \code{NetworkX} to create a graph from the point cloud and to identify the largest connected component (LCC). We present the plot of $d_\mathrm{max}$ versus the number of retrograde star particles within the corresponding LCC in the right panel of \figref{fig:kmeans_dmax_retro}. We minimize the residual of fitting a $\Gamma$ shape to the plot following \citet{2025CNSNS.14908923S} to locate the knee (indicated by the pink dashed line), which corresponds to the $d_\mathrm{max}$ delimiting the bar ($d_\mathrm{max} = 0.018$). Bar particles are those within the LCC corresponding to the limiting $d_\mathrm{max}$. We obtain \numprint{133417} bar particles in the K11 simulation at $t = 0.075$~Gyr, as shown in the bottom right panel of \figref{fig:bar_classification}.

\subsection{Comparison with the action-based bar classification method}
\label{sec:bar_class2}

\begin{figure}
  \resizebox{\hsize}{!}{\includegraphics{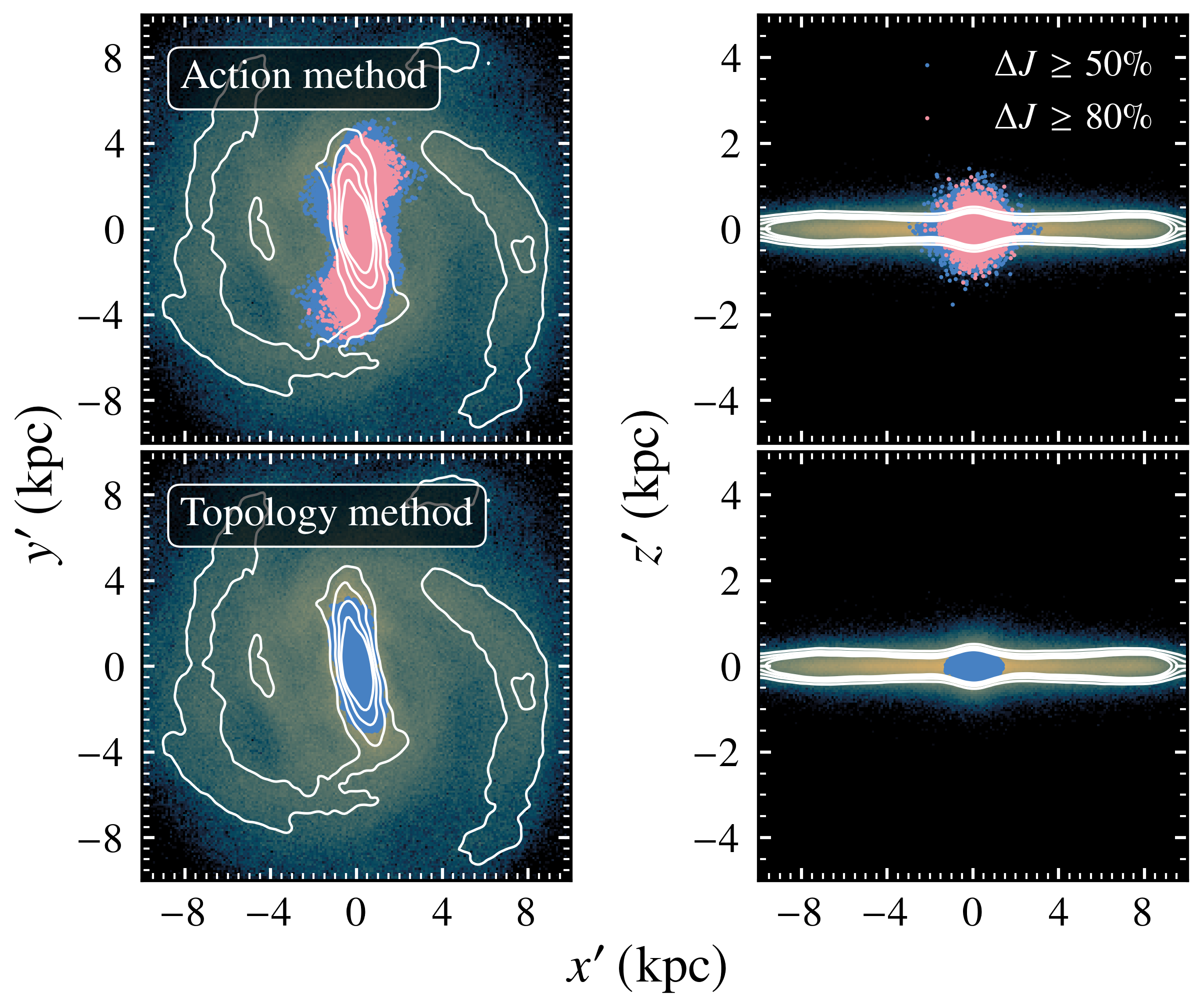}}
  \caption{Top panels: Face-on and side-on view of the $G_\mathrm{LMC}$, overplotted by bar particles obtained using the action-based method (blue and pink scatter points) in the isolated K10 simulation at $t = 0.075$~Gyr. Bottom panels: the $G_\mathrm{LMC}$ in the same snapshot, overplotted by bar particles obtained using the topology-based method (blue scatter points.)}
  \label{fig:method_compare}
\end{figure}

\begin{figure}
  \resizebox{\hsize}{!}{\includegraphics{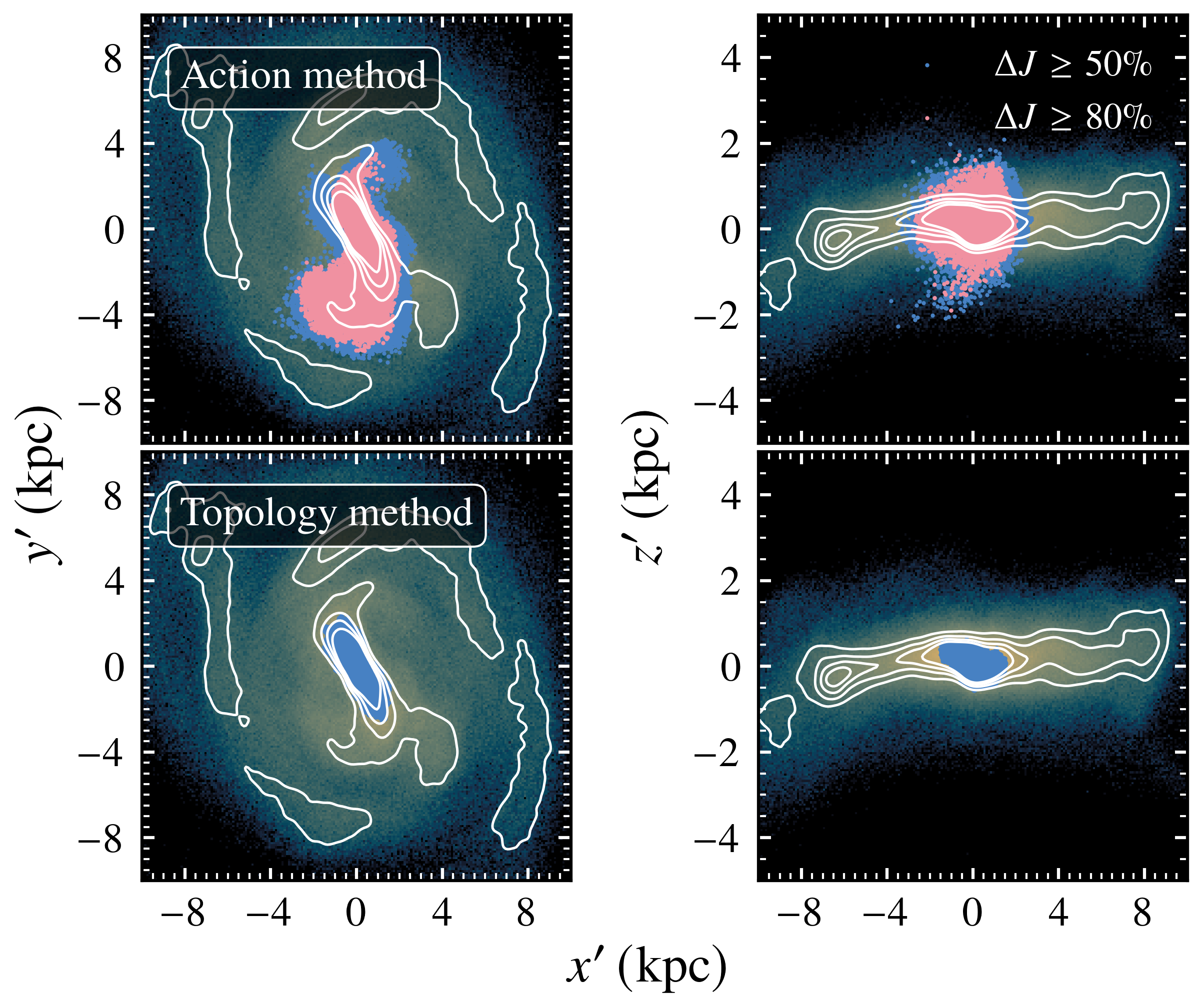}}
  \caption{Top panels: Face-on and side-on view of the $G_\mathrm{LMC}$, overplotted by bar particles obtained using the action-based method (blue and pink scatter points) in the K11 simulation at $t = 0.075$~Gyr (second pericentric passage of the $G_\mathrm{SMC}$). Bottom panels: the $G_\mathrm{LMC}$ in the same snapshot, overplotted by bar particles obtained using the topology-based method (blue scatter points.)}
  \label{fig:K11_method_compare}
\end{figure}

Here, we present the results of extracting bar particles using the action conservation method \citep{2026MNRAS.548ag611I}. We compute the radial ($J_R$), vertical ($J_z$), and azimuthal actions ($J_{\phi}$) of the $G_\mathrm{LMC}$ star particles using the
St\"{a}ckel fudge method \citep{10.1111/j.1365-2966.2012.21757.x} implemented in \code{agama} \citep{2019MNRAS.482.1525V}. Note that \citet{2026MNRAS.548ag611I} mainly used \code{GALPY} for their analysis, but they checked that using either package only produces slightly different results. We approximate the potential of both the dark matter halo and stellar component using multipole expansions. The total action is calculated by
\begin{align*}
    J = \sqrt{J_R^2 + J_z^2 + J_{\phi}^2} \,.
\end{align*}
Bar particles are those with high variation in total action within a short time interval $\Delta t$.

As an example, we apply both methods to the snapshot at $t = 0.075$~Gyr in the isolated K10 simulation, where the disk remains flat throughout the simulation, and disk particles have well-conserved total action. \figrefalt{fig:method_compare} shows the $G_\mathrm{LMC}$ bar particles obtained using the action-based method (top panels) and the topology-based method (bottom panels, as discussed in Sect.~\ref{sec:method}). In the top panels, the (pink) blue scatter points denote the star particles with a change in total action ($\Delta J \geq 80\%$) $\Delta J \geq 50\%$ within $\Delta t = 75$~Myr. In the bottom panels, the blue scatter points are the bar particles we obtain using the topology-based method. \figrefalt{fig:method_compare} shows that the bar particles obtained using our topology-based method are much better constrained. The differences between the bar particles obtained by the two methods could be due to the relatively low cadence of the KRATOS simulations ($\Delta t = 75$~Myr) compared to $\Delta t = 10$~Myr in \citet{2026MNRAS.548ag611I}，see their Fig.~1.

In addition, the $G_\mathrm{LMC}$ disk in our interacting models could be strongly perturbed by the $G_\mathrm{SMC}$ pericentric passages, and disk star particles generally do not have well-conserved total action. \figrefalt{fig:K11_method_compare} shows the $G_\mathrm{LMC}$ bar particles obtained using the action-based method (top panels) and the topology-based method (bottom panels) in the interacting K11 simulation at $t = 0.075$~Gyr (second $G_\mathrm{SMC}$ pericentric passage). The plot demonstrates that, when the disk is strongly perturbed by tidal interactions, the action conservation method cannot cleanly separate the bar and disk particles, while the topology-based method still performs well. Therefore, we adopt the topology-based method in our analysis.

\section{Alternative alignment method}
\label{sec:ori_align}
In this section, we present an alternative alignment method used for comparison.

We follow the initial centering and alignment process as in Sect.~\ref{sec:method}. The aligned galaxy corresponds to the middle panel of \figref{fig:center_align}. We then align the $G_\mathrm{LMC}$ with the 
$G_\mathrm{LMC}$ disk within $5$~kpc, without re-centering the $G_\mathrm{LMC}$. We iteratively obtain bar particles and realign the $G_\mathrm{LMC}$, until the number of obtained bar particles converge within $1\%$. This is the final set of bar particles. We then align the $G_\mathrm{LMC}$ with the principal direction of the annulus with radius $R_0 < R < 5$~kpc, where $R_0$ is the bar radius.

\begin{figure}
  \resizebox{\hsize}{!}{\includegraphics{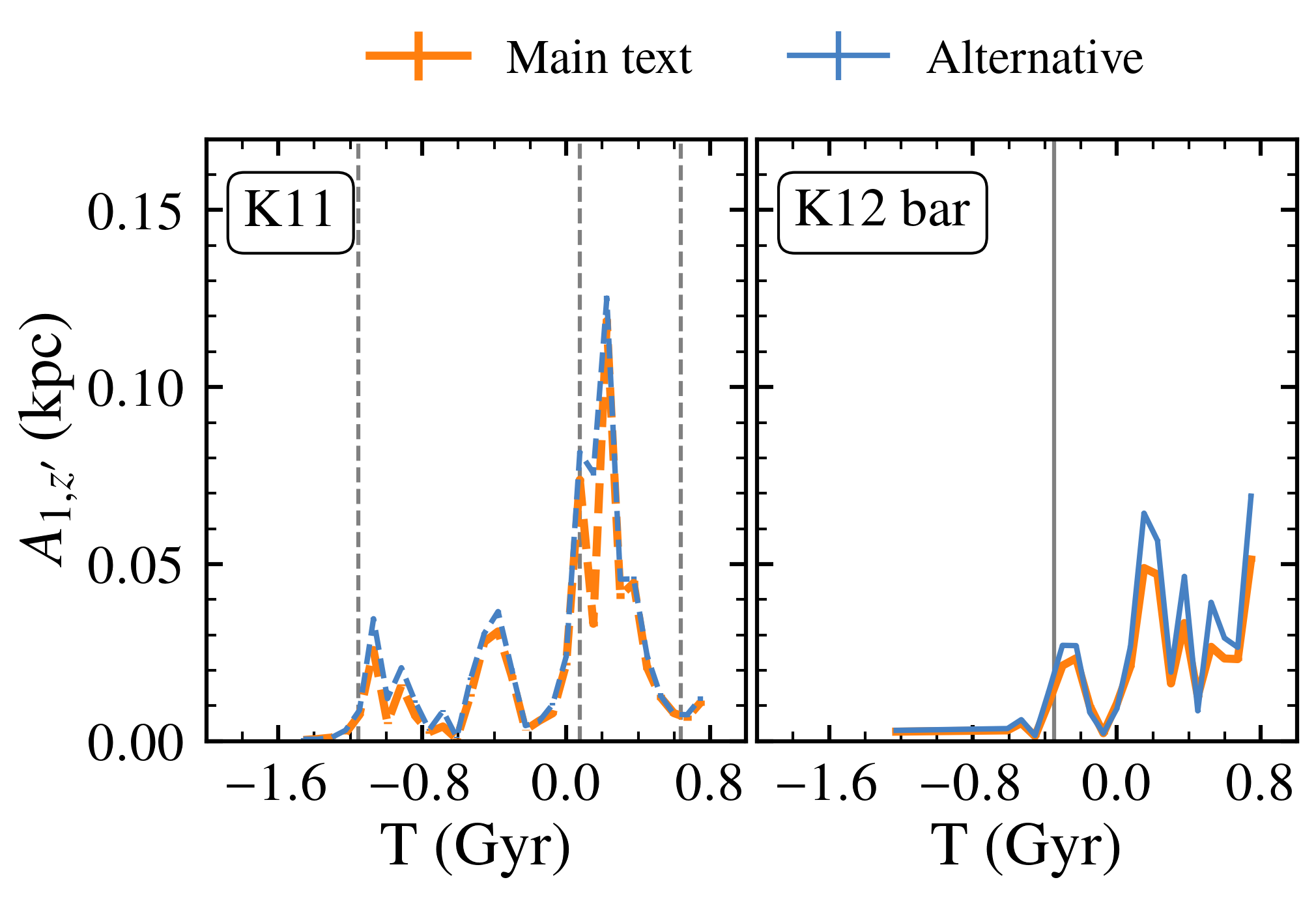}}
  \caption{Time evolution of the $m = 1$ vertical position asymmetry in the $G_\mathrm{LMC}$ bar in the simulations K11 and K12 for the alignment method in the main text (orange) and the alternative alignment method (blue).}
  \label{fig:t_A1_zvz_bar_compare_align}
\end{figure}

\begin{figure}
  \resizebox{\hsize}{!}{\includegraphics{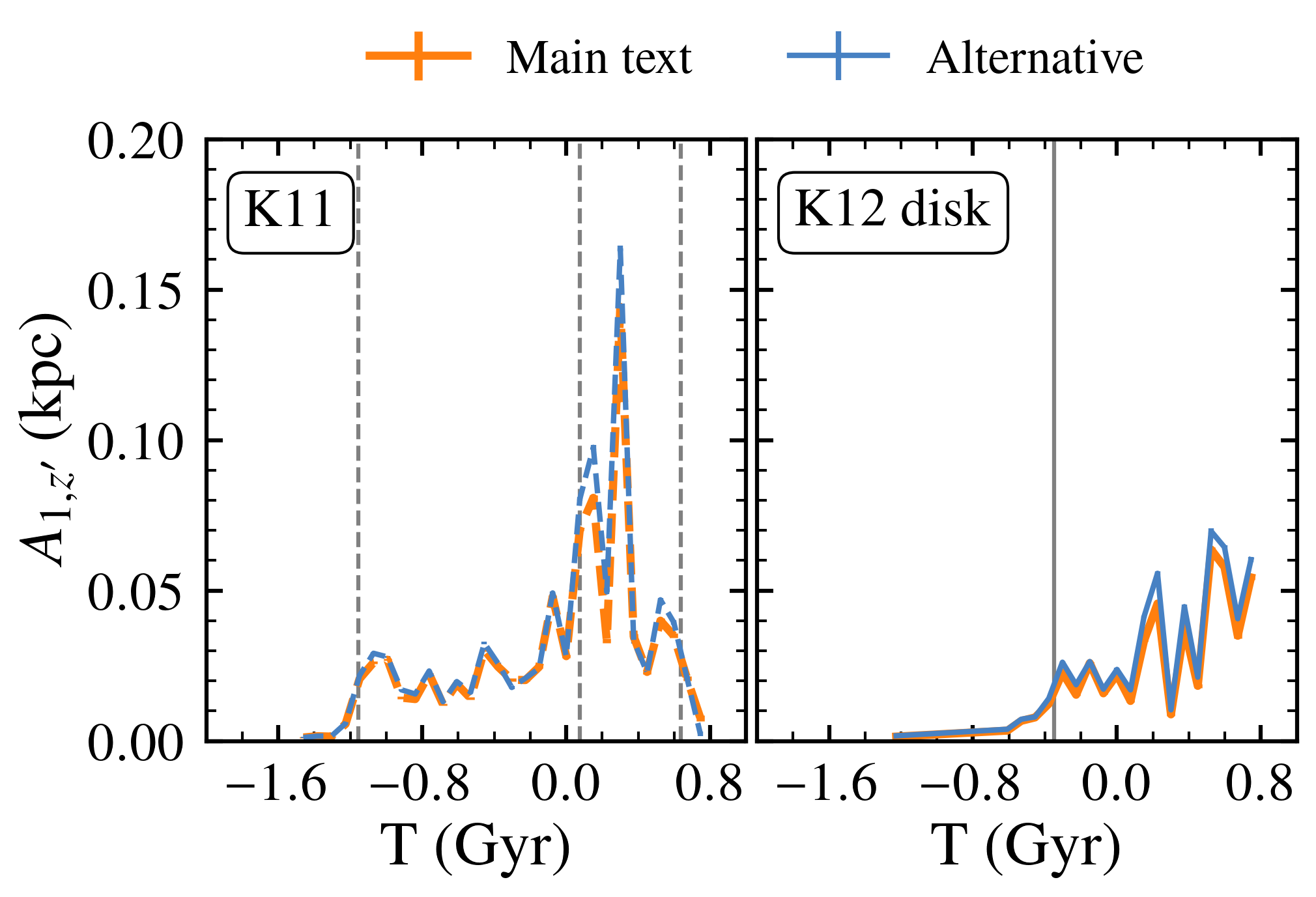}}
  \caption{Time evolution of the $m = 1$ vertical position asymmetry in the $G_\mathrm{LMC}$ disk in the simulations K11 and K12 for the alignment method in the main text (orange) and the alternative alignment method (blue).}
  \label{fig:t_A1_zvz_disk_compare_align}
\end{figure}

For comparison, we provide the time evolution of the vertical position asymmetry in the $G_\mathrm{LMC}$ bar and disk in the simulations K11 and K12 for both alignment methods in \figsref{fig:t_A1_zvz_bar_compare_align} and \ref{fig:t_A1_zvz_disk_compare_align}. The plots show that the alignment method in the main text slightly reduces the $m = 1$ vertical asymmetry in the $G_\mathrm{LMC}$ bar and disk. We verified that changing the alignment method does not affect the $m = 2$ vertical asymmetry in the $G_\mathrm{LMC}$ bar and disk. In particular, the signatures of the $G_\mathrm{SMC}$ pericentric passages are largely preserved, irrespective of the adopted alignment method.

\section{Supplementary figures}
\label{sec:supplementary}
\begin{figure}
  \resizebox{\hsize}{!}{\includegraphics{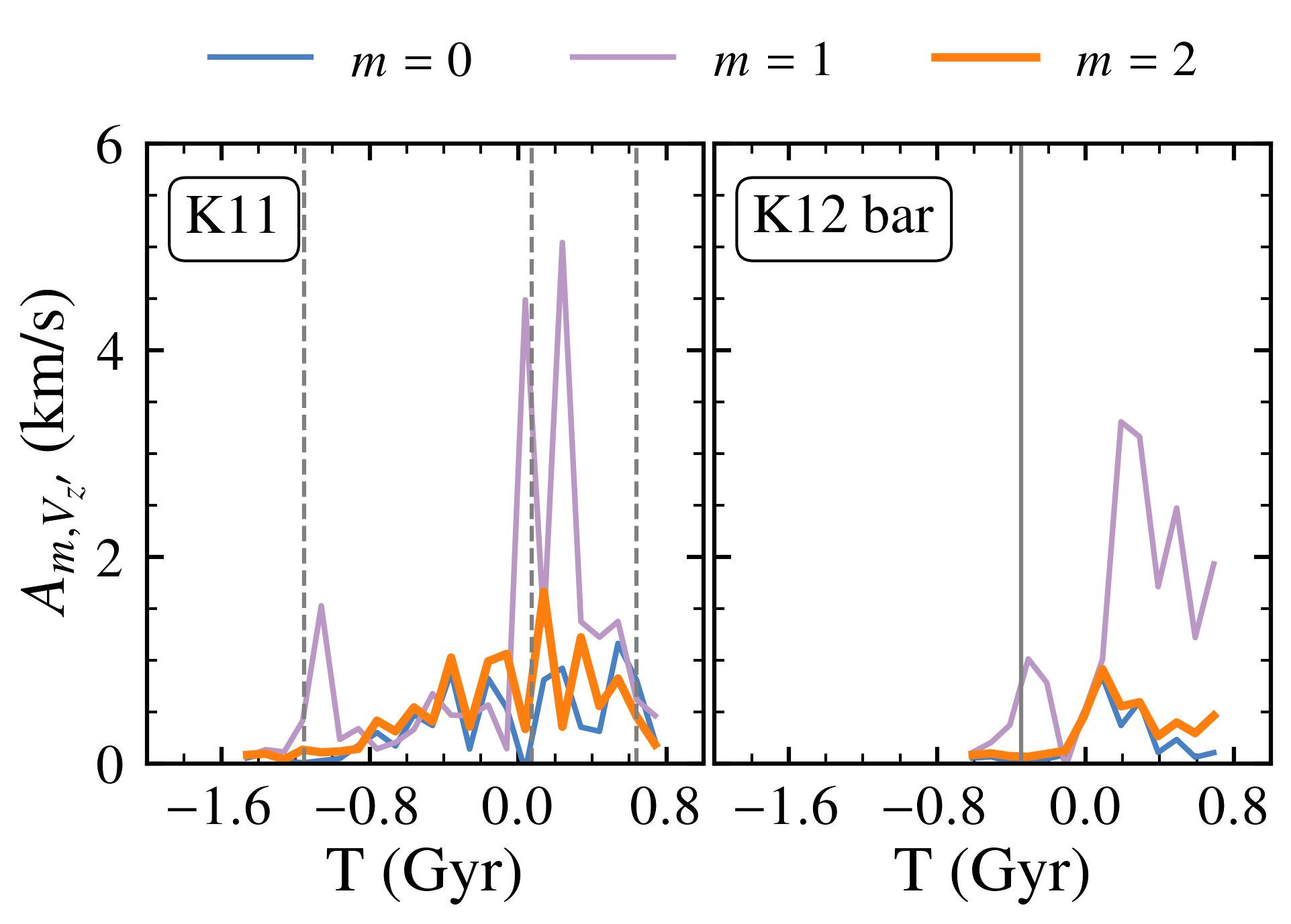}}
  \caption{Time evolution of the $m = 0, 1, 2$ $V_{z^{\prime}}$ asymmetry in the $G_\mathrm{LMC}$ bar in the K11 and K12 simulations.}
  \label{fig:t_A1_vz_bar_A1_A2}
\end{figure}

\figrefalt{fig:t_A1_vz_bar_A1_A2} shows the time evolution of the $m=0, 1, 2$ vertical velocity asymmetry in the $G_\mathrm{LMC}$ bar in the simulation set K11 and K12. The $m = 0$ vertical asymmetry closely follows the $m = 2$ vertical asymmetry in the $G_\mathrm{LMC}$ bar. This is due to Fourier decomposing a long $G_\mathrm{LMC}$ bar, where the majority of the bar particles have the same (or opposite) azimuthal angles.

\begin{figure}
  \resizebox{\hsize}{!}{\includegraphics{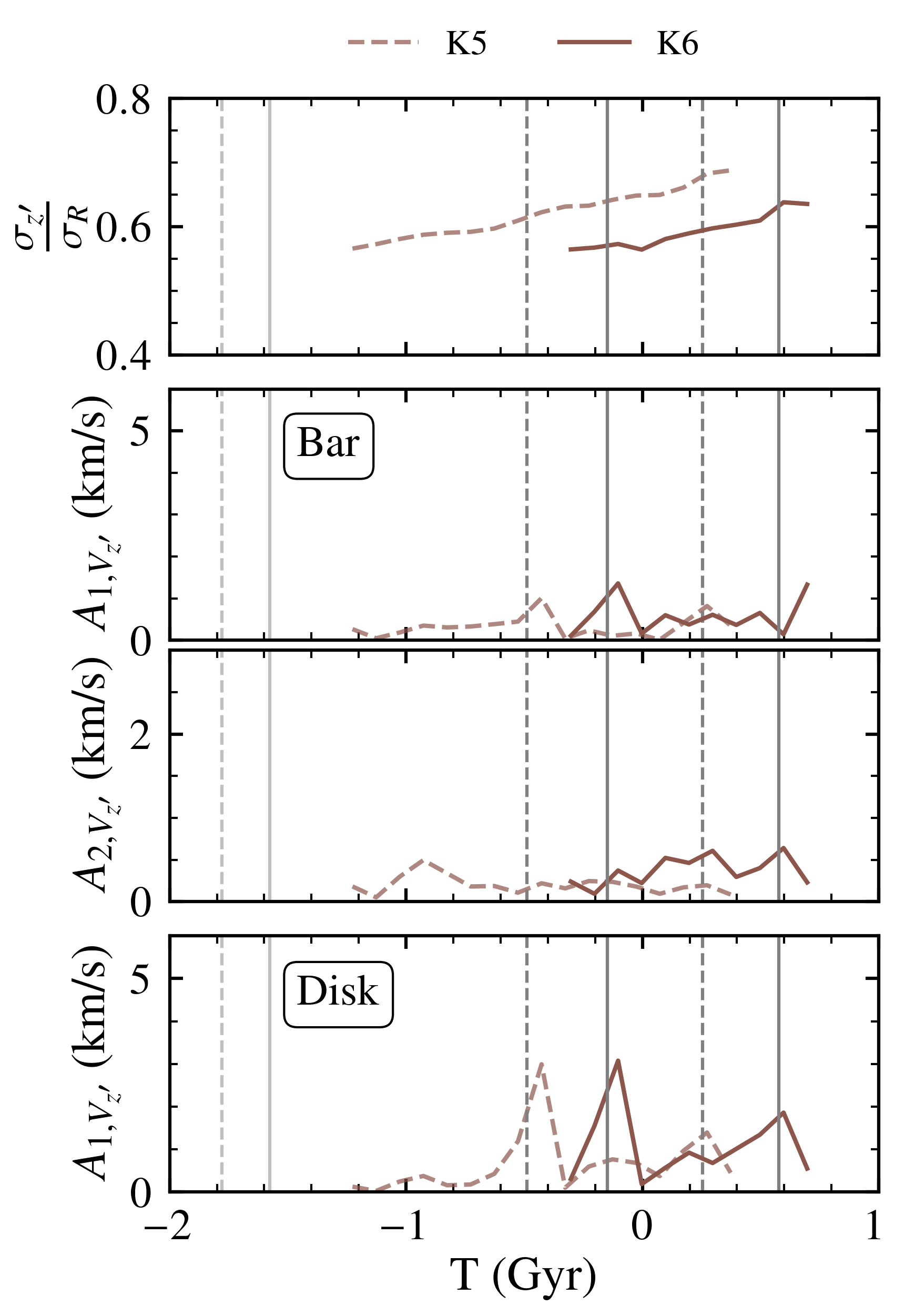}}
  \caption{From top to bottom: time evolution of the velocity dispersion ratio in the inner $G_\mathrm{LMC}$ ($R\leq 2$~kpc), $m = 1$ $V_{z^{\prime}}$ asymmetry in the $G_\mathrm{LMC}$ bar, $m = 2$ $V_{z^{\prime}}$ asymmetry in the $G_\mathrm{LMC}$ bar and $m = 1$ $V_{z^{\prime}}$ asymmetry in the $G_\mathrm{LMC}$ disk in the simulations K5 and K6. The vertical grey dashed (solid) lines correspond to the pericenters of the $G_\mathrm{LMC}+G_\mathrm{SMC}$ ($G_\mathrm{LMC}+G_\mathrm{SMC} + G_\mathrm{MW}$) models. }  \label{fig:t_A1_zvz_K4_bar_disk_smoothed}
\end{figure}

\begin{figure}
  \resizebox{\hsize}{!}{\includegraphics{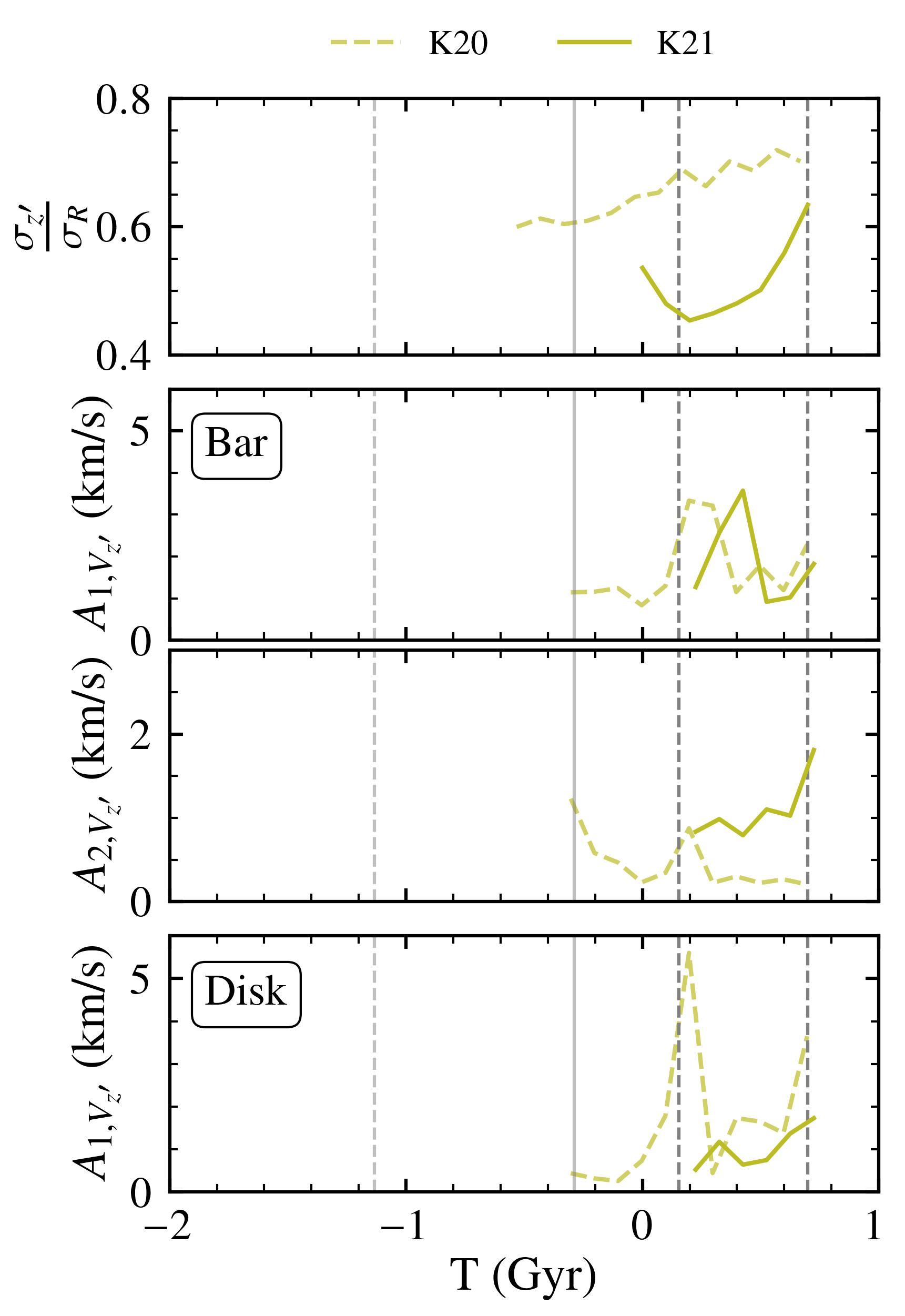}}
  \caption{From top to bottom: time evolution of the velocity dispersion ratio in the inner $G_\mathrm{LMC}$ ($R\leq 2$~kpc), $m = 1$ $V_{z^{\prime}}$ asymmetry in the $G_\mathrm{LMC}$ bar, $m = 2$ $V_{z^{\prime}}$ asymmetry in the $G_\mathrm{LMC}$ bar and $m = 1$ $V_{z^{\prime}}$ asymmetry in the $G_\mathrm{LMC}$ disk in the simulations K20 and K21. The vertical grey dashed (solid) lines correspond to the pericenters of the $G_\mathrm{LMC}+G_\mathrm{SMC}$ ($G_\mathrm{LMC}+G_\mathrm{SMC} + G_\mathrm{MW}$) models.}  \label{fig:t_A1_zvz_K19_bar_disk_smoothed}
\end{figure}

\figsrefalt{fig:t_A1_zvz_K4_bar_disk_smoothed} and \ref{fig:t_A1_zvz_K19_bar_disk_smoothed} present the results of Fourier decomposing the vertical velocities of the $G_\mathrm{LMC}$ bar and disk for the remaining four simulations, K5 ($G_\mathrm{LMC}+G_\mathrm{SMC}$), K6 ($G_\mathrm{LMC}+G_\mathrm{SMC} + G_\mathrm{MW}$), K20 ($G_\mathrm{LMC}+G_\mathrm{SMC}$) and K21 ($G_\mathrm{LMC}+G_\mathrm{SMC} + G_\mathrm{MW}$). The top panels in \figsref{fig:t_A1_zvz_K4_bar_disk_smoothed} and \ref{fig:t_A1_zvz_K19_bar_disk_smoothed} show that the $G_\mathrm{LMC}$ bar in K21 buckles after the first $G_\mathrm{SMC}$ passage, corresponding to a sudden rise in the velocity dispersion ratio and an increase in $A_{2,V_{z^{\prime}}}$. The $G_\mathrm{LMC}$ bar in K5, K6 and K20 does not buckle throughout the simulation. Due to the $G_\mathrm{LMC}$ bar not being fully formed, we remove the first three snapshots in K20 and K21 in the bottom three panels of \figref{fig:t_A1_zvz_K19_bar_disk_smoothed}. \figsrefalt{fig:t_A1_zvz_K4_bar_disk_smoothed} and \ref{fig:t_A1_zvz_K19_bar_disk_smoothed} demonstrate that each $G_\mathrm{SMC}$ passage induces vertical asymmetries in both the $G_\mathrm{LMC}$ bar and disk, consistent with the results in Sect.~\ref{sec:result}.

\begin{figure}
  \resizebox{\hsize}{!}{\includegraphics{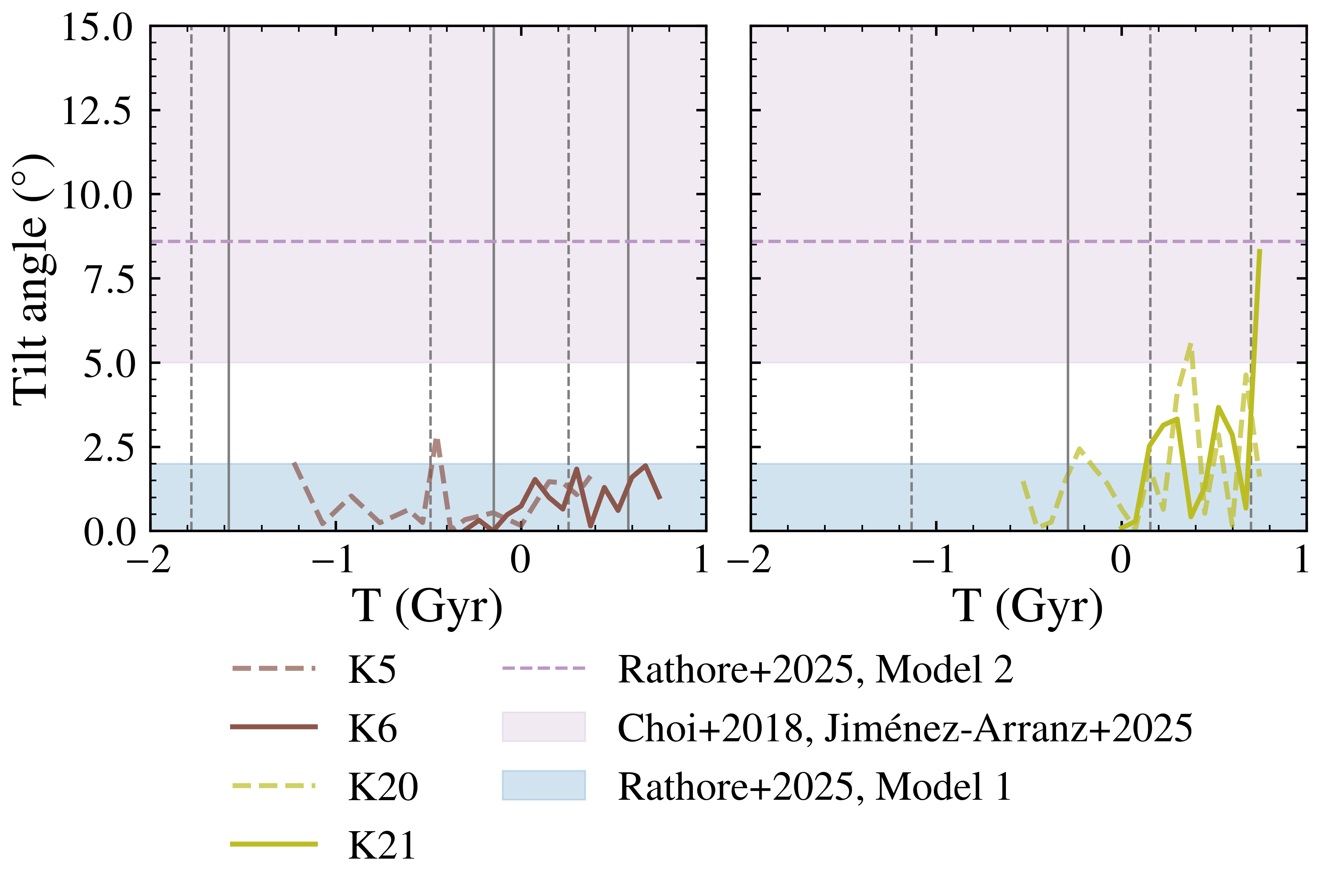}}
  \caption{Time evolution of the tilt of the $G_\mathrm{LMC}$ bar in the simulations K5, K6, K20 and K21. The purple shaded region shows the estimated tilt of the $G_\mathrm{LMC}$ bar in \citet{2018ApJ...866...90C} and \citet{2025A&A...698A..88J}. The purple dashed line denotes the tilt angle of the $G_\mathrm{LMC}$ bar at the present time ($\sim 8.6\degree$) in Model 2 of \citet{2025ApJ...988...79R}. The blue shaded region shows the $1\sigma$ spread around the mean tilt angle of the $G_\mathrm{LMC}$ bar ($\sim 1\degree$) in Model 1 of \citet{2025ApJ...988...79R}.}
  \label{fig:t_tilt_K4_K19}
\end{figure}

\figrefalt{fig:t_tilt_K4_K19} illustrates the time evolution of the estimated tilt angle of the $G_\mathrm{LMC}$ bar in the four simulations. The plot shows that the tilt of the $G_\mathrm{LMC}$ bar in the these four simulation models is not observationally significant. In particular, the tilt in the models with the lowest mass ratio between the $G_\mathrm{LMC}$ and the $G_\mathrm{SMC}$, K5 and K6, remains within the $1\sigma$ spread around the mean tilt angle of the $G_\mathrm{LMC}$ bar in Model 1 throughout the simulation. Although the tilt of the $G_\mathrm{LMC}$ bar at the end of the simulation K21 is close to that in Model 2, this is due to the bar being deformed by the strong buckling, which biases the estimation of the tilt angle.

\section{Simple model of a buckling bar}
\label{sec:buckling_model}
In this section, we construct a simple model of a buckling bar. We first consider \numprint{1000} particles on nearly elliptical orbits, undergoing harmonic oscillations ($x$ is aligned with the semi-major axis of the ellipse).
The phase of each particle can be written as $\theta_x = \omega t + P$, where $\omega$ is the in-plane frequency of the particle, and $P$ is the initial azimuth in the $x$-$y$ plane, added for randomization. Therefore, the in-plane motion of the particle can be written as $x =  R \cos(\theta_x)$, $y = c R \sin(\theta_x)$, where $R$ is the length of the semi-major axis of the ellipse and $c$ is the ratio between the semi-minor and semi-major axis of the ellipse. We set $c = 0.2$, which forms the long bar in our model. To simulate the upward or downward-curved banana shape of a buckling bar, we relate the vertical position of the particle, $z$, to $x$ by $z \propto x^2$.
When buckling, the vertical frequency $\nu$ and the in-plane frequency $\omega$ are in $2:1$ resonance \citep[e.g.,][]{2019A&A...629A..52L}. Therefore, $\nu = 2\omega$, and $z = c x^2 \cos(\nu t)$.

\begin{figure}
  \resizebox{\hsize}{!}{\includegraphics{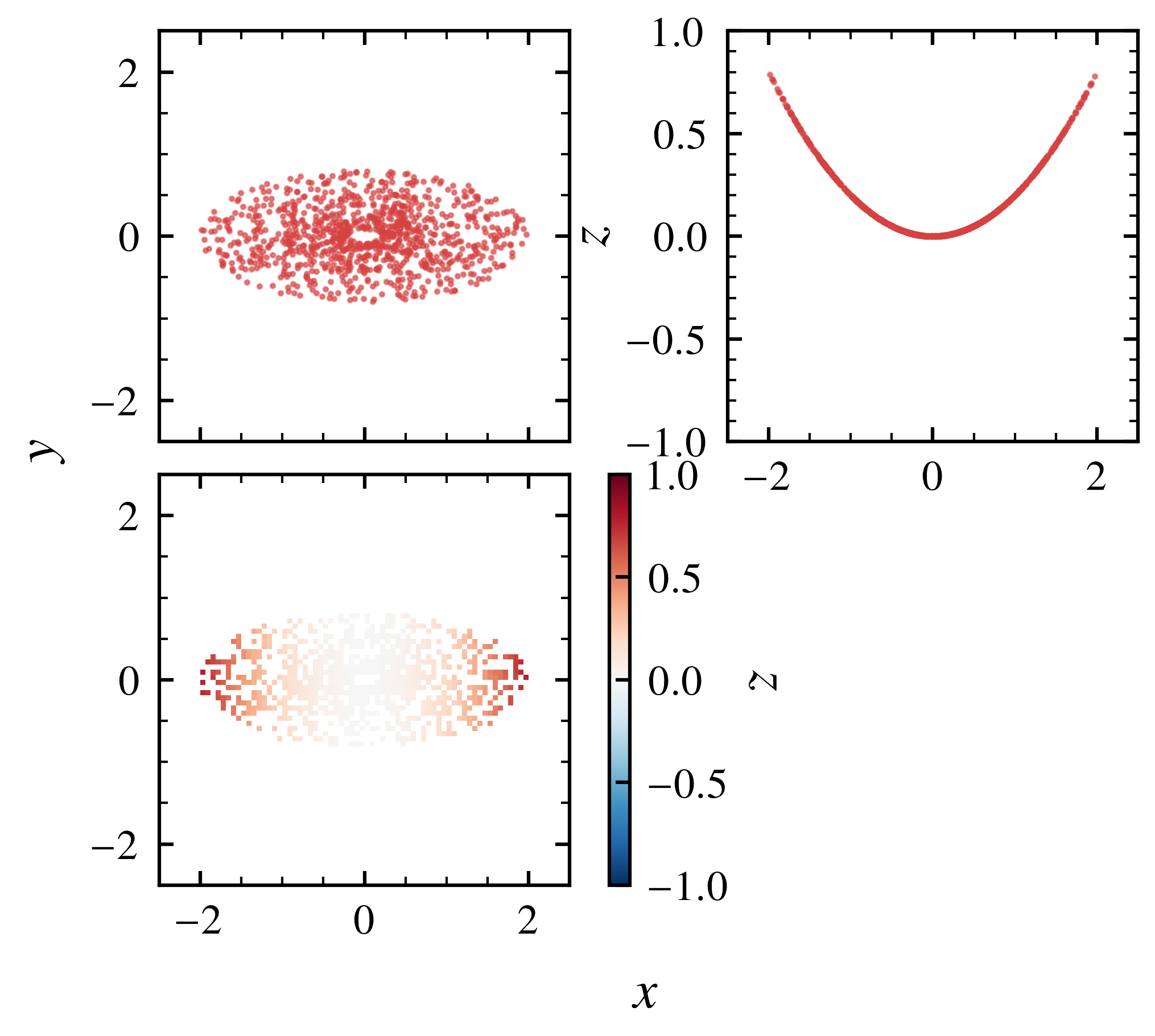}}
  \caption{Face-on (top left) and side-on (top right) view, as well as the face-on average vertical velocity map (bottom left) of the buckling bar particles at $t = \pi$.}
  \label{fig:buckling_model}
\end{figure}

We evolve these particles for $t = 2\pi$. The face-on density map (top left) and side-on density map (top right) of the buckling bar at a representative snapshot $t = \pi$ is shown in \figref{fig:buckling_model}. The top right panel shows the characteristic upward-curved banana shape. 

As shown in the left panel of \figref{fig:decomp_buckling_tilting_model}, we perform Fourier decomposition on the vertical positions of these bar stars. The plot demonstrates that when Fourier decomposing the vertical positions where the majority of the particles have the same (or opposite) azimuthal angles, the resulting $A_{0,z}$ and $A_{2,z^\prime}$ have nearly the same trend with time. Note that for this simple model, $A_{1,z^{\prime}}$ is nonzero.

\begin{figure}
  \resizebox{\hsize}{!}{\includegraphics{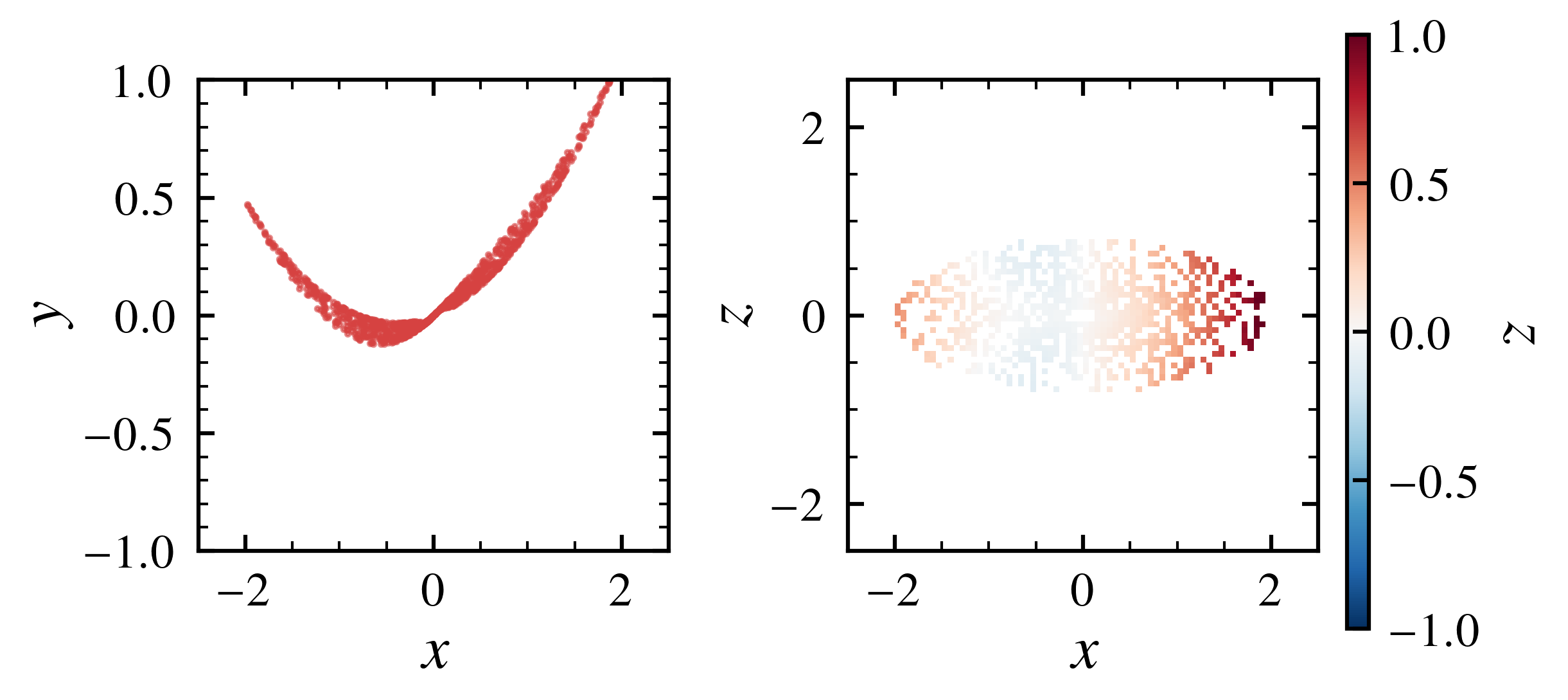}}
  \caption{Side-on (left) view and the face-on average vertical velocity map (right) of the buckling and tilting bar particles at $t = \pi$.}
  \label{fig:buckling_tilting_model}
\end{figure}

\begin{figure}
\centering
  \resizebox{\hsize}{!}{\includegraphics{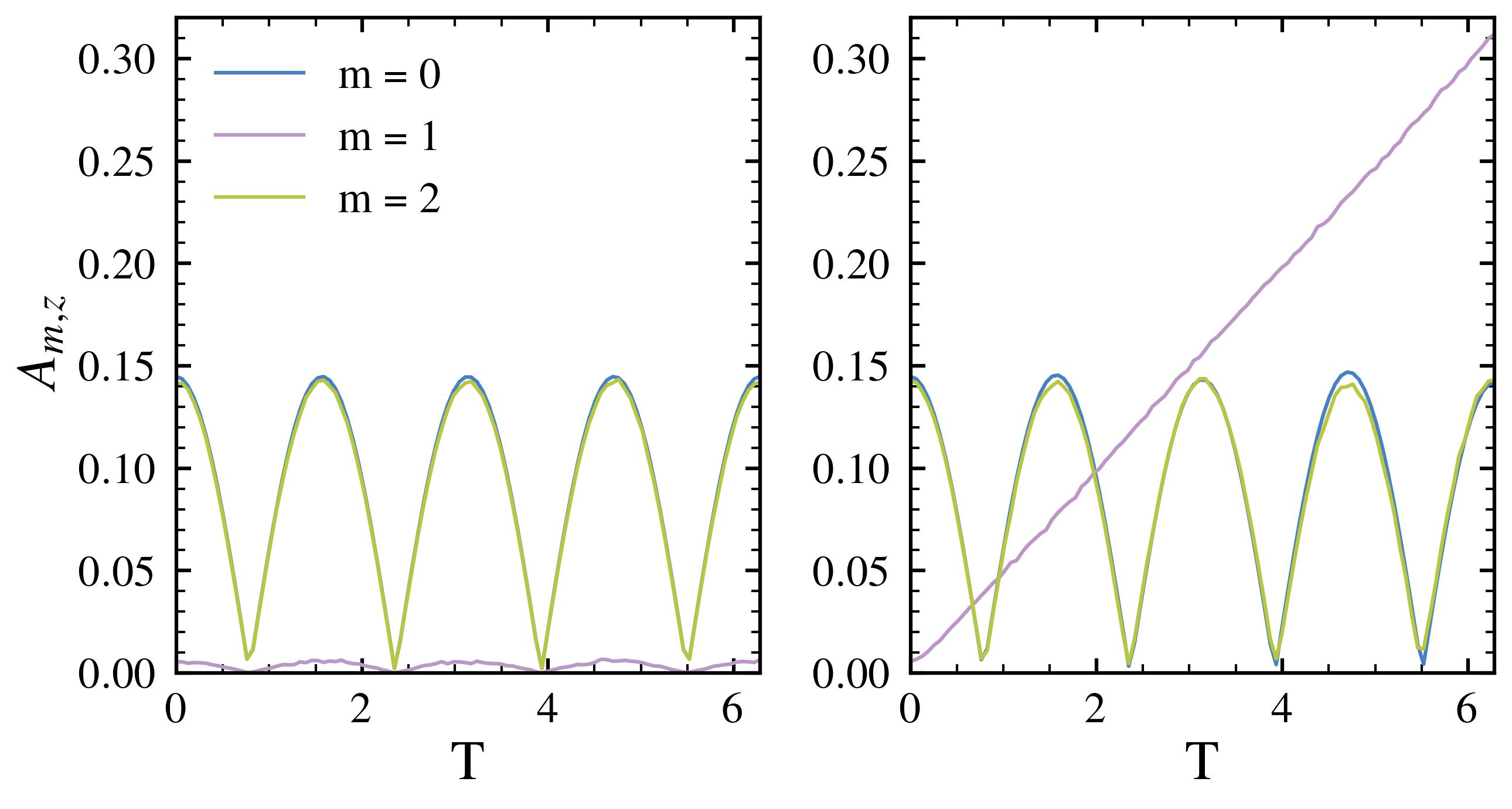}}
  \caption{Fourier decomposition of the vertical position of the buckling particles (left) and the buckling and tilting bar particles (right) for $m = 0\,,1\,,2$. The particles are evolved for $t = 2\pi$.}
  \label{fig:decomp_buckling_tilting_model}
\end{figure}

In order to simulate an $m = 1$ perturbation to the buckling bar, we add a ``tilting" component with a constant rate $\omega_t$ to the vertical position $z$ of the particle. Now, $z = R\omega_t t  \cos(\theta_x) + c x^2 \cos(\nu t)$. We set $\omega_t = 0.1$. We show the buckling and tilting bar at $t = \pi$ in \figref{fig:buckling_tilting_model}. Note that the left right panel in \figref{fig:buckling_tilting_model} qualitatively resembles the fifth subplot of the left panels in \figref{fig:K11_bar_disk}, where the buckling bar appears to be lopsided.

The Fourier decomposition results are shown in the right panel of \figref{fig:decomp_buckling_tilting_model}. As the buckling bar becomes increasingly lopsided, $A_{1, z^\prime}$ linearly increases with time. The Fourier decomposition method separate the buckling ($m = 0$ and $m = 2$) and the tilting mode ($m = 1$) well.

\onecolumn
\end{appendix}

\end{CJK*}
\end{document}